\documentclass[12pt,a4paper,reqno]{amsart}

\usepackage{amssymb}
\usepackage{amsmath}
\usepackage{latexsym}
\usepackage{exscale}
\usepackage[latin1]{inputenc}
\usepackage{epsfig}
\usepackage{verbatim}
\usepackage{graphicx}
\usepackage{subfigure}

\newcommand{\R}{\mathbb R}
\newcommand{\C}{\mathbb{C}}

\newcommand{\N}{\mathbb{N}}

\newcommand{\la}{\lambda}

\newcommand{\al}{\alpha}

\newcommand{\G}{\Gamma}

\newcommand{\de}{\delta}

\newcommand{\norm}[1]{\left\Vert#1\right\Vert}
\newcommand{\abs}[1]{\left| #1 \right|}
\newcommand{\ip}[2]{\langle #1 , #2 \rangle}

\newcommand{\E}[1]{\mathbb{E}\left\{#1\right\}}
\newcommand{\Prb}[1]{\mathbb{P}\left\{#1\right\}}

\numberwithin{equation}{section} 

\newtheorem{thm}{Theorem}[section] 
\newtheorem{cor}[thm]{Corollary}

\newtheorem{defi}[thm]{Definition}
\newtheorem{rem}[thm]{Remark}

\newtheorem{lem}[thm]{Lemma}

\begin{document}

\title[GREEDY TYPE ALGORITHMS FOR RIP MATRICES]{GREEDY TYPE ALGORITHMS FOR RIP MATRICES. A STUDY OF TWO SELECTION RULES.}

\author{Eugenio Hernández}

\address{Eugenio Hernández
\\
Departamento de Matem\'aticas
\\
Universidad Aut\'onoma de Ma\-drid
\\
28049 Madrid, Spain}

\email{eugenio.hernandez@uam.es}

\author{Daniel Vera}

\address{Daniel Vera
\\
Departamento de Matem\'aticas
\\
Universidad Aut\'onoma de Madrid
\\
28049 Madrid, Spain}

\email{daniel.vera@uam.es}

\thanks{Research supported by Grant  MTM2010-16518 of Spain.}

\date{\today}
\subjclass[2000]{41A46, 68W20}

\keywords{Compressed sensing, convergence rate, greedy algorithms,
random matrices, restricted isometry property.}


\maketitle

\begin{abstract}
On \cite{NeVe1} some consequences of the Restricted Isometry
Property (RIP) of matrices have been applied to develop a greedy
algorithm called ``ROMP" (Regularized Orthogonal Matching Pursuit)
to recover sparse signals and to approximate non-sparse ones. These
consequences were subsequently applied to other greedy and
thresholding algorithms like ``SThresh", ``CoSaMP", ``StOMP" and
``SWCGP". In this paper, we find another consequence of the RIP
property and use it to analyze the approximation to $k$-sparse
signals with \textit{Stagewise Weak} versions of Gradient Pursuit
(SWGP),
 Matching Pursuit (SWMP) and Orthogonal Matching Pursuit
(SWOMP) algorithms described in in \cite{BlDa2}.
We combine the above mentioned
algorithms with another selection rule similar to the ones that appeared in \cite{CoDaDe2}  and  \cite{DTDS}
showing that results are obtained with less restrictions in the RIP constant, but we need a smaller threshold parameter for
the coefficients. The results of some experiments are shown.

\end{abstract}

\section{Introduction}\label{Intro}
One problem in Compressed Sensing (CS) is to reconstruct a
$k-$sparse vector (all except $k$ elements are zero)
$\mathbf{x}\in\R^N$ from a lower dimension vector
$\mathbf{y}=\mathbf{\Phi} \mathbf{x}$, where
$\mathbf{\Phi}\in\R^{m\times N}$ is called a CS matrix or
measurements ensemble. The aim of CS is to compress the signal while
taking samples at the same time in such a way that a ``good"
reconstruction is possible. The next definition (see
\cite{CanTao2}) is a sufficient condition for the so called CS
matrices to yield exact reconstruction of sparse signals with the
Basis Pursuit (BP) algorithm or $\ell^1$ minimization with equality
constraints (as firstly proposed by Donoho and collaborators for
dictionaries in the signal processing community). This property will
play a central role in the results developed here.
\begin{defi}\label{d:RIP_deltaK}
Given $k\in \N$, a matrix $\mathbf{\Phi}\in\C^{m\times N}$ ($m>k$)
is said to satisfy the Restricted Isometry Property with parameter
$\de_k$, $0 < \delta_k < 1$ (called the Restricted Isometry
Constant), if
\begin{equation}\label{e:RIP_deltaK}
    (1-\de_k)\norm{\mathbf{x}}^2_{\ell^2(\R^N)} \leq \norm{\mathbf{\Phi} \mathbf{x}}^2_{\ell^2(\R^m)} \leq
    (1+\de_k)\norm{\mathbf{x}}^2_{\ell^2(\R^N)}
\end{equation}
for \textbf{all} $k-$sparse vectors $\mathbf{x}\in\R^N$.
\end{defi}

It is known that those matrices that satisfy RIP and allow the least
number of measurements $m$ for reconstruction of all sparse signals
are some random matrices (therefore reconstruction is in
probability). For example, in \cite{CanTao2} and \cite{Don}, it is
shown that it suffices to take $m$ linearly with the sparsity $k$
and polylogarithmic with the ambient dimension $N$, \emph{i.e.}
$m\geq C k \log (N/k)$ for Gaussian and Bernoulli matrices, and
$m\geq C k \log^6 (N)$ for Fourier matrices (for better bounds see
\cite{RuVe1}) to reconstruct, with high probability, a $k$-sparse
vector with the BP algorithm.

\

It has become standard to use greedy algorithms to
iteratively identify the support $\Gamma^\sharp:=\text{supp
}(\mathbf{x})$ of a sparse signal. This is done by computing the
inner product of the residue $\mathbf{r}^{n-1}$ of the approximation
at step $n-1$ and the columns of the matrix $\mathbf{\Phi}$,
\emph{i.e.} $\mathbf{g}^n=\mathbf{\Phi}^\ast\mathbf{r}^{n-1}$ ($\mathbf{\Phi}^\ast $ denotes de transpose of $\mathbf{\Phi}$) , and
then select the largest element(s) (in absolute value) in
$\mathbf{g}^n=(g_1^n,\ldots,g_N^n)^t$, where each
$g_i^n=\ip{\phi_i}{\mathbf{r}^{n-1}}$ and $\phi_i$ is the $i$-th
column of $\mathbf{\Phi}$. Since $\mathbf{\Phi}$ verifies the RIP
property, then such an inner product gives an idea on where the
support may be because the square of the energies of the $k$-sparse
vector signal $\mathbf{x}$ and the observation vector
$\mathbf{y}=\mathbf{\Phi}\mathbf{x}$ should not differ more than
$\delta_k$. To see this more clearly, suppose we know the true
support of $\mathbf{x}$, $\Gamma^\sharp=\text{supp }(\mathbf{x})$,
and $\mathbf{y}=\mathbf{\Phi}\mathbf{x}$ is the observation. Writing
$\mathbf{\Phi}_\Gamma$ (resp. $\mathbf{x}_\Gamma$) to denote the
matrix $\mathbf{\Phi}$ (resp. the vector $\mathbf{x}$) restricted to
the columns (resp. the elements) indexed by $\Gamma\subset
\{1,2,\ldots,N\}$, and $(\mathbf{\Phi}_{\Gamma^\sharp})^\dag
=(\mathbf{\Phi}^\ast_{\Gamma^\sharp}\mathbf{\Phi}_{\Gamma^\sharp})^{-1}\mathbf{\Phi}^\ast_{\Gamma^\sharp}$
for the pseudo inverse of $\mathbf{\Phi}_{\Gamma^\sharp}$ (which
exists by (\ref{e:RIP_deltaK})), we can recover $\mathbf{x}$ from
$\mathbf{y}$ using $(\mathbf{\Phi}_{\Gamma^\sharp})^\dag$ since
$\mathbf{y}=\mathbf{\Phi}\mathbf{x}=\mathbf{\Phi}_{\Gamma^\sharp}\mathbf{x}_{\Gamma^\sharp}$
and
\begin{equation} \label{reconstruction}
(\mathbf{\Phi}_{\Gamma^\sharp})^\dag \mathbf{y}=
(\mathbf{\Phi}^\ast_{\Gamma^\sharp}\mathbf{\Phi}_{\Gamma^\sharp})^{-1}\mathbf{\Phi}^\ast_{\Gamma^\sharp}
    \mathbf{\Phi}_{\Gamma^\sharp}\mathbf{x}_{\Gamma^\sharp}=\mathbf{x}_{\Gamma^\sharp}.
\end{equation}
\,

We shall use the notation $\mathbb{R}^\Gamma$ to denote the subspace
of the ambient space $\mathbb{R}^N$ with significant coordinates in
$\Gamma\subset \{1,\ldots,N\}$. Notice the prominent role of
$\mathbf{\Phi}^\ast_{\Gamma^\sharp}:\mathbb{R}^m\rightarrow\mathbb{R}^{\Gamma^\sharp}$
in the above argument. The matrix $\mathbf{\Phi}^\ast$ (the transpose of $\mathbf{\Phi}$) will be used
in the algorithms below. Also, $\mathbf{r}\in \text{span
}(\mathbf{\Phi}_\Gamma)$ means that $\mathbf{r}$ is a linear
combination of the columns of $\mathbf{\Phi}$ indexed by $\Gamma$.

\

Section \ref{Sec:SupId_RIP} studies conditions to identify the
support of a sparse signal sensed with RIP matrices using different
greedy type algorithms. A review of \textit{Stagewise Weak} versions of the Gradient
Pursuit (SWGP), Matching Pursuit (SWMP), and Orthogonal Matching Pursuit
(SWOMP)  algorithms, as first proposed in \cite{BlDa2}, is done in Subsection \ref{Sec:grdy_algthms}.
The main novelty, as pointed out in \cite{BlDa2},
is that not only one but several elements are allowed to be
selected in each iteration. This is a feature also present in \cite{DTDS}, a paper published
in 2012 but circulated since 2006 as a preprint. Weak algorithms were used in non-linear approximation
theory before the development of Compressed Sensing (see \cite{Jon87}, \cite{Bar93} and more
recently \cite{DT96}, \cite{Tem00}, \cite{Tem03}, \cite{Tem11} and the references therein), but in all of these
works only one element was selected at each iteration.

\

 Known properties
of RIP matrices are stated in Subsection \ref{Sec:RIP_cnsq}, while a
new property of these matrices is proved in Subsection
\ref{Sec:One_more} (see Lemma \ref{l:PHIstar_below}). We give
conditions on $\mathbf{\Phi}$ and on the weakness parameter $\alpha$ of the selection rule to identify the support of a sparse
signal in Subsection \ref{Sec:Select_on_Weak_GPMPOMP} (see Theorem
\ref{t:ERC_Weak_PHIstar_below}) for all of the \textit{Stagewise Weak} algorithms mentioned above.

\

 Another selection rule, called
\textbf{relaxed} in this paper, is introduced in Section
\ref{Sec:Select_on_Rlx_Weak_GPMPOMP} given rise to new algorithms that we name Relaxed Weak
Gradient Pursuit (RWGP), Relaxed Weak Matching Pursuit (RWMP) and Relaxed Weak Orthogonal Matching
Pursuit (RWOMP). The strategy of selecting several elements in each iteration is
also used in these algorithms. In the Relaxed selection rule, elements are chosen if their
magnitude is larger than a fraction of the energy of the residue at an iteration. This procedure
has also been used in \cite{CoDaDe2}  and  \cite{DTDS} (see the second paragraph in Section \ref{Sec:Select_on_Rlx_Weak_GPMPOMP}). The name
Weak Relaxed has appeared in the non-linear approximation theory associated to greedy algorithms (see
\cite{DT96}, \cite{Tem00}), but with a different meaning that in this paper.  In Theorem
\ref{t:ERCNew_Weak_PHIstar_below} we give conditions on a matrix $\mathbf{\Phi}$ satisfying RIP and on the weakness parameter $\alpha$ of
the Relaxed selection rule to identify the support of a sparse signal.

\

In Section \ref{Sec:Convrgnc} the convergence of all the above
algorithms is studied. The energy of the residual of the observation
at iteration $n$ is compared with the energy at iteration $n-1$,
i.e, $ \|\mathbf{r}^{n}\|_{\ell^2} \leq C_k
\|\mathbf{r}^{n-1}\|_{\ell^2}\,.$ For the GP, SWGP and RWGP
algorithms we establish in Theorem \ref{t:Convrg_GP_PHIstar_below}
the above inequality with
$C_k=(1-\frac{1-\delta_k}{k(1+\delta_k)})^{1/2}< 1$; this is a more
explicit version than the ones already known for GP and SWGP (see
details in Section \ref{Sec:Convrgnc}). For the SWMP, SWOMP, RWMP,
and RWOMP the result is stated in Theorem
\ref{t:Convrg_WMP_PHIstar_below} giving $C_k' =
(1-\frac{(1-\delta_k)^2}{k})^{1/2}<1$.
Notice that Gradient Pursuit algorithms seem to have stronger rate
of convergence than Matching Pursuit ones, a fact present in the
experiments shown for images in Section \ref{Sec:Exprmnts}.

\

In Section \ref{sS:CS_RndMtx} we study the behavior of the selection
rules with particular Gaussian and Bernoulli random matrices not
necessarily satisfying RIP. Here we prove that with high probability
the algorithms allow to recover the position of the $k$ entries of a
given $k$-sparse signal. Finally, some experiments are shown in
Section \ref{Sec:Exprmnts} for the algorithms described in the above
sections, and compare results with already existing algorithms to
recover sparse signals and approximate compressible images with a
sparse representation.


\section{Support Identification with RIP.}\label{Sec:SupId_RIP}
We will review Stagewise Weak versions of the Gradient
Pursuit (SWGP), Matching Pursuit (SWMP) and Orthogonal Matching Pursuit
(SWOMP) algorithms and some consequences of the RIP property. Then, we
will develop a new consequence of RIP and find some conditions so
the algorithms select elements on the support of a $k-$sparse
signal $\mathbf{x}$ on each iteration. 

\

\subsection{SWGP, SWMP and SWOMP
algorithms}\label{Sec:grdy_algthms} The Stagewise Weak Gradient
Pursuit (SWGP), Stagewise Weak Matching Pursuit (SWMP) and Stagewise
Weak Orthogonal Matching Pursuit (SWOMP) algorithms select a set of
elements (possibly not new) in each iteration by comparing the
maximum of the inner products between the columns of the measurement
ensemble with the residue at the previous iteration. This stage in
the algorithms is the selection rule, where a weakness parameter
$\alpha$ is introduced. The main differences between these
algorithms are the direction search and the way to update the
approximation. For some history on MP and OMP see \cite{Tr1} and
references therein. The \emph{stagewise weak selection rule} is
defined in \cite{BlDa2}.

\vskip 0.9cm
\begin{center}
STAGEWISE WEAK MATCHING PURSUIT (SWMP)\\
\end{center}
The MP algorithm was the first of the greedy algorithms to be used
in signal processing (see \cite{MaZh93}); as far as we know, the weakness parameter appeared for
the first time in \cite{Jon87} and has been extensively used in non-linear approximation
(see \cite{DT96}, \cite{Tem00}, \cite{Tem03}, \cite{Tem11} and the references therein).

In the initialisation we set:
at iteration $n=0$ the residue is $\mathbf{r}^0=\mathbf{y}$, the
approximation to the observation is $\mathbf{y}^0=0$ and the
estimation of the signal is $\mathbf{x}^0=0$. Recall that $\phi_i$
denotes the $i$-th column of $\mathbf{\Phi}$. The loop until some
criteria are met follows the next steps:
\begin{itemize}
    \item $\mathbf{g}^n=\mathbf{\Phi}^\ast \mathbf{r}^{n-1}$, the ``proxy" of the signal.
    \item $\mathcal{I}_n := \mathcal{I}_n(\alpha) :=\left\{ i: \abs{g^n_i} \ge \al \norm{\mathbf{\Phi}^\ast
    \mathbf{r}^{n-1}}_{\ell^\infty(\R^N)}\right\}$, the stagewise weak selection rule with $0<\alpha\leq 1$.
    \item $\mathbf{y}^n=\mathbf{y}^{n-1}+\sum_{i\in\mathcal{I}_n} g^n_i \mathbf{\phi}_i
        =\mathbf{y}^{n-1}+\sum_{i\in\mathcal{I}_n} \ip{\mathbf{\phi}_i}{\mathbf{r}^{n-1}} \mathbf{\phi}_i$,
    approximation to the observation.
    \item $x^n_i=x^{n-1}_i+g^n_i=x^{n-1}_i+\ip{\mathbf{\phi}_i}{\mathbf{r}^{n-1}}$, $i\in\mathcal{I}_n$
        and $x_j^n = x_j^{n-1}$ if $j\notin \mathcal{I}_n$, estimation of the signal.
    \item $\mathbf{r}^n=\mathbf{r}^{n-1}-\sum_{i\in\mathcal{I}_n} g^n_i \mathbf{\phi}_i
            =\mathbf{r}^{n-1}-\sum_{i\in\mathcal{I}_n} \ip{\mathbf{\phi}_i}{\mathbf{r}^{n-1}} \mathbf{\phi}_i$, the residual.
\end{itemize}
By recursion one can see from the definition of $\mathbf{y}^n$ and
$\mathbf{r}^n$ that
$\mathbf{r}^n=\mathbf{r}^{n-1}-(\mathbf{y}^n-\mathbf{y}^{n-1})=\cdots=\mathbf{y}-\mathbf{y}^n$.

\vskip 0.3cm
\begin{center}
STAGEWISE WEAK ORTHOGONAL MATCHING PURSUIT (SWOMP)\\
\end{center}
SWOMP is similar to SWMP. Once $\mathcal{I}_n$ has been selected in
SWMP, the updating of the approximation with
$\sum_{i\in\mathcal{I}_n} g^n_i \mathbf{\phi}_i$ might not be the
best approximation from the subspace spanned by the columns
$\{\mathbf{\phi}_i\}_{i\in\mathcal{I}_n}$. For SWOMP instead, the
update of the approximation is $\mathbf{y}^n=P_{\G^n}\mathbf{y}$
where $P_{\G^n}$ is the orthogonal projection into the span of the
columns $\G^n$ of $\mathbf{\Phi}$:
$\mathbf{y}^n=\mathbf{\Phi}_{\G^n}\mathbf{\Phi}^\dagger_{\G^n}\mathbf{y}
= \mathbf{\Phi}_{\G^n}(\mathbf{\Phi}_{\G^n}^\ast
\mathbf{\Phi}_{\G^n})^{-1}\mathbf{\Phi}_{\G^n}^\ast \mathbf{y}$ on
the indices $\G^n=\bigcup_{k=1}^n \mathcal{I}_k$ and therefore the
residue becomes $\mathbf{r}^n=\mathbf{y}-P_{\G^n}\mathbf{y}$. The
residue is then orthogonal to all elements previously selected as
can be seen from

\begin{center}\begin{tabular}{r @{} l}
$\mathbf{\Phi}_{\G^n}^\ast \mathbf{r}^n$
    & $=\mathbf{\Phi}_{\G^n}^\ast(\mathbf{y}-\mathbf{\Phi}_{\G^n}\mathbf{\Phi}_{\G^n}^\dagger \mathbf{y})$\\
$$  & $= \mathbf{\Phi}_{\G^n}^\ast \mathbf{y} -
        \mathbf{\Phi}_{\G^n}^\ast\mathbf{\Phi}_{\G^n}(\mathbf{\Phi}_{\G^n}^\ast \mathbf{\Phi}_{\G^n})^{-1} \mathbf{\Phi}_{\G^n}^\ast \mathbf{y}=0.$\\
\end{tabular}\end{center}

Thus, at every iteration new elements are selected. The
initialisation is as in SWMP and the recursion loop is:
\begin{itemize}
    \item $\mathbf{g}^n=\mathbf{\Phi}^\ast \mathbf{r}^{n-1}$, the ``proxy" of the signal.
    \item $\mathcal{I}_n := \mathcal{I}_n (\alpha) :=\left\{ i: \abs{g^n_i} \ge \al \norm{\mathbf{\Phi}^\ast
        \mathbf{r}^{n-1}}_{\ell^\infty(\R^N)}\right\}$, the stagewise weak selection rule with $0< \alpha \leq 1.$
    \item $\G^n=\G^{n-1}\bigcup \mathcal{I}_n$, update of
        selected elements.
    \item $\mathbf{x}^n:=\mathbf{x}^n|_{\G^n}=\mathbf{\Phi}^\dagger_{\G^n}\mathbf{y}$, estimation of the signal.
    \item $\mathbf{y}^n= \mathbf{\Phi}_{\G^n}\mathbf{\Phi}^\dagger_{\G^n}\mathbf{y}$,
        approximation to the observation.
    \item $\mathbf{r}^n=\mathbf{y}-\mathbf{y}^n$, the residual.
\end{itemize}

\vskip 0.3cm
\begin{center}
STAGEWISE WEAK GRADIENT PURSUIT (SWGP)\\
\end{center}

The Gradient Pursuit (GP) is described in \cite{BlDa1} and the Stagewise Weak Gradient Pursuit (SWGP)
was developed in \cite{BlDa2} (together with some other variants). We describe SWGP since GP is obtained from SWGP by setting the weakness parameter
$\alpha =1$ in the selection rule.

 At iteration $n=0$ we have:
$\mathbf{x}^0=0$, the estimation of the signal;
$\mathbf{r}^0=\mathbf{y}$, the residue; $\G^0=\emptyset$, the
support. Then, the recursion at step $n$ until some criteria are met
is the following:
    \begin{itemize}
        \item $\mathbf{g}^n=\mathbf{\Phi}^\ast \mathbf{r}^{n-1}$, the ``proxy" of the signal.
        \item $\mathcal{I}_n := \mathcal{I}_n(\alpha) :=
                \{i:\abs{g^n_i}\geq\al\norm{\mathbf{r}^{n-1}}_{\ell^\infty(\R^N)}\}$, the stagewise weak
                selection rule with $0 < \alpha \leq 1.$
        \item $\G^n=\G^{n-1}\bigcup \mathcal{I}_n$, the updated support.
        \item $\mathbf{d}^n_{\Gamma^n}=\mathbf{\Phi}_{\Gamma^n}^\ast \mathbf{r}^{n-1}$, the updated direction.
        \item $a^n=\frac{\ip{\mathbf{r}^{n-1}}{\mathbf{\Phi}_{\Gamma^n} \mathbf{d}_{\Gamma^n}^n}}
            {\norm{\mathbf{\Phi}_{\Gamma^n}
            \mathbf{d}_{\Gamma^n}^n}_{\ell^2(\mathbb{R}^m)}^2}$, the
            optimised step.
        \item $\mathbf{x}^n:=\mathbf{x}^n_{\G^n}=\mathbf{x}^{n-1} +
            a^n\mathbf{d}^n_{\G^n}$, the
            estimation to the signal.
        \item $\mathbf{y}^n=\mathbf{y}^{n-1}
            + a^n\mathbf{\Phi}_{\Gamma^n}\mathbf{d}^n_{\Gamma^n}$, the
            approximation to the observation.
        \item $\mathbf{r}^n=\mathbf{r}^{n-1}-a^n\mathbf{\Phi}_{\Gamma^n}\mathbf{d}^n_{\Gamma^n}$.
    \end{itemize}
Again, one can prove that $\mathbf{r}^n=\mathbf{y}-\mathbf{y}^n$.

In  \cite{BlDa2} it is proved that
SWGP converges to the solution at least as good as the simpler
version GP.

\

\subsection{Some Consequences of RIP}\label{Sec:RIP_cnsq} This
section is based on some implicit results in \cite{CanTao2}, see
especially Lemma 2.1, that were made explicit in \cite{NeVe1}; for
the proofs the reader can also see \cite{CoDaDe2}.

\begin{lem}\label{l:CONSQ_RIP}
Assume that $\mathbf{\Phi}\in\R^{m\times N}$ satisfies RIP with
$\de_k$, $\abs{\G}\leq k$, $\text{supp}(\mathbf{u})= \G$. Then

    \begin{equation}\label{e:PHIstar_above}
            \norm{\mathbf{\Phi}_\G}_{\ell^2(\R^\G)\to\ell^2(\R^m)}
            = \norm{\mathbf{\Phi}^\ast_\G}_{\ell^2(\R^m)\to\ell^2(\R^\G)}
            \leq (1+\de_k)^{1/2},
            \end{equation}
    \begin{equation}\label{e:PHIstarPHI}
            (1-\de_k)\norm{\mathbf{u}}_{\ell^2(\R^\G)}
            \leq \norm{\mathbf{\Phi}^\ast_\G\mathbf{\Phi}_\G \mathbf{u}}_{\ell^2(\R^\G)}
            \leq (1+\de_k)\norm{\mathbf{u}}_{\ell^2(\R^\G)},
            \end{equation}
    \begin{equation}\label{e:PHIstarPHI1}
            (1+\de_k)^{-1}\norm{\mathbf{u}}_{\ell^2(\R^\G)}
            \leq \norm{(\mathbf{\Phi}^\ast_\G\mathbf{\Phi}_\G)^{-1} \mathbf{u}}_{\ell^2(\R^\G)}
            \leq (1-\de_k)^{-1}\norm{\mathbf{u}}_{\ell^2(\R^\G)},
            \end{equation}\vskip 0.01cm
    \text{For disjoint sets $\G', \G$ such that} $\abs{\G'\bigcup\G}\leq
    k$, we have
            \begin{equation}\label{e:Almost_orth}
            \norm{\mathbf{\Phi}^\ast_{\G'}\mathbf{\Phi}_\G \mathbf{u}}_{\ell^2(\R^{\G'})} \leq
            \de_k\norm{\mathbf{u}}_{\ell^2(\R^\G)}.
            \end{equation}\vskip 0.01cm

\end{lem}
All of these results have become standard for the greedy algorithms
in compressed sensing.

\

\subsection{One more consequence}\label{Sec:One_more} We derive
here another bound as a consequence of the RIP property.

\begin{lem}\label{l:PHIstar_below}
Let $\mathbf{\Phi}$ verify RIP with parameter $\de_k$ and let $
\abs{\Gamma} \leq k$. For all $\mathbf{r}\in\text{span
}(\mathbf{\Phi}_\Gamma)$
    \begin{equation}\label{e:PHIstar_below}
            \norm{\mathbf{\Phi}^\ast_\Gamma \mathbf{r}}_{\ell^2(\R^\Gamma)} \ge (1-\de_k)^{½}
            \norm{\mathbf{r}}_{\ell^2(\R^m)}
    \end{equation}
\end{lem}
\textbf{\emph{Proof.}} For the pseudo-inverse we have
\begin{center}\begin{tabular}{r @{} l}
$\norm{\mathbf{\Phi}^\dagger_\Gamma\mathbf{r}}_{\ell^2(\R^\Gamma)}$
    & $=\norm{(\mathbf{\Phi}^\ast_\Gamma \mathbf{\Phi}_\Gamma)^{-1}\mathbf{\Phi}^\ast_\Gamma \mathbf{r}}_{\ell^2(\R^\Gamma)}$\\

$$  & $\leq (1-\de_k)^{-1} \norm{\mathbf{\Phi}^\ast_\Gamma \mathbf{r}}_{\ell^2(\R^\Gamma)},$\\
\end{tabular}\end{center}
where we have used (\ref{e:PHIstarPHI1}) in the last inequality. We
next use this result and the Schwarz inequality to get
\begin{eqnarray*} \norm{\mathbf{\Phi}^\ast_\Gamma \mathbf{r}}^2_{\ell^2(\R^\Gamma)}
&\geq &  (1-\de_k)\norm{\mathbf{\Phi}^\dagger_\Gamma \mathbf{r}}_{\ell^2(\R^\Gamma)} \norm{\mathbf{\Phi}^\ast_\Gamma \mathbf{r}}_{\ell^2(\R^\Gamma)}  \\
& \geq & (1-\de_k) \ip{\mathbf{\Phi}^\dagger_\Gamma \mathbf{r}}{\mathbf{\Phi}^\ast_\Gamma \mathbf{r}}
        = (1-\de_k) (\mathbf{\Phi}^\dagger_\Gamma \mathbf{r})^\ast \mathbf{\Phi}^\ast_\Gamma \mathbf{r} \\
& = & (1-\de_k) \mathbf{r}^\ast \mathbf{\Phi}_\Gamma (\mathbf{\Phi}^\ast_\Gamma \mathbf{\Phi}_\Gamma)^{-1} \mathbf{\Phi}^\ast_\Gamma \mathbf{r} \\
& = &   (1-\de_k) \mathbf{y}^\ast_\Gamma \mathbf{\Phi}^\ast_\Gamma \mathbf{\Phi}_\Gamma (\mathbf{\Phi}^\ast_\Gamma \mathbf{\Phi}_\Gamma)^{-1}
        \mathbf{\Phi}^\ast_\Gamma \mathbf{\Phi}_\Gamma \mathbf{y}_\Gamma \\
& = & (1-\de_k) \mathbf{y}^\ast_\Gamma \mathbf{\Phi}^\ast_\Gamma \mathbf{\Phi}_\Gamma  \mathbf{y}_\Gamma
= (1-\de_k) \norm{\mathbf{r}}^2_{\ell^2(\R^m)},
\end{eqnarray*}
using the fact that, since $\mathbf{r}$ is in the span of
$\mathbf{\Phi}_\Gamma$, we may write
$\mathbf{r}=\mathbf{\Phi}_\Gamma \mathbf{y}_\Gamma$ for some
$\mathbf{y}_\Gamma$, and conclude the proof.
\hfill $\blacksquare$ \vskip .5cm   

\subsection{Support Identification with SWGP, SWMP and
SWOMP}\label{Sec:Select_on_Weak_GPMPOMP} We now give sufficient
conditions on the matrix $\mathbf{\Phi}$ so that the SWGP, SWMP and
SWOMP algorithms select elements on the support $\G^\sharp$ of the
sparse signal $\mathbf{x}$. The SWOMP algorithm will select new atoms
in each iteration due to the orthogonality with previous residues,
therefore convergence to exact reconstruction of $k-$sparse vectors
is guaranteed in at most $k$ iterations when the condition is met.
As previously mentioned, the stagewise weak selection rule is
\begin{equation}\label{e:Sel_Rule}
    \mathcal{I}_n := \mathcal{I}_n(\alpha) := \left\{ i: \abs{g^n_i} \ge \al \norm{\mathbf{\Phi}^\ast \mathbf{r}^{n-1}}_{\ell^\infty(\R^N)}\right\},
\end{equation}
for some $\alpha\in(0,1]$. For the next result we follow the line of reasoning of
\cite{Tr1}, (see also \cite{GrVa} and \cite{BlDa2}), where the results are given for OMP on \emph{quasi-incoherent dictionaries}.
Here we use Lemma \ref{l:PHIstar_below}.

\begin{thm}\label{t:ERC_Weak_PHIstar_below}
Let $\mathbf{\Phi}$ satisfies RIP with $\de_{k+1}$. A sufficient
condition for the SWGP, SWMP and SWOMP algorithms with selection rule $\mathcal{I}_n(\alpha)$ given by
(\ref{e:Sel_Rule}), $0 <\al \leq 1$, to identify elements in the support $\G^\sharp$
of the $k-$sparse signal $\mathbf{x}$ is
\begin{equation}\label{e:cond_selecGP}
\al > \frac{\sqrt{k}\de_{k+1}}{1-\de_k}.
\end{equation}
\end{thm}
\textbf{\emph{Proof.}} Since $\text{supp}(\mathbf{x})=\G^\sharp$,
then $\mathbf{r}^0\in\text{span}(\mathbf{\Phi}_{\G^\sharp})$. The
algorithms update the approximations and residuals precisely in the
indices contained in $\mathcal{I}_n$, so that to proceed by induction we can assume that after
$n-1$ iterations we have
$\mathbf{r}^{n-1}\in\text{span}(\mathbf{\Phi}_{\G^\sharp})$. We drop
the superindex of $\mathbf{r}$ for the proof. The condition
$\mathcal{I}_n \subset \Gamma^\sharp$ is implied by
$$\norm{\mathbf{\Phi}_{\G^{\sharp c}}^\ast \mathbf{r}}_{\ell^\infty(\R^{\G^{\sharp c}})} < \al \norm{\mathbf{\Phi}^\ast_{\G^\sharp} \mathbf{r}}_{\ell^\infty(\R^{\G^\sharp})}.$$
Using the assumption on $\mathbf{r}$, we can write
$\mathbf{r}=\mathbf{\Phi}_{\G^\sharp} \mathbf{y}_{\G^\sharp}$ for
some $\mathbf{y}_{\G^\sharp}$. Rearranging to express the condition
as a quotient (called the \emph{greedy selection ratio}), squaring
and choosing $\la$ as one index in $\G^{\sharp c}$ with the largest
value, it yields

\begin{eqnarray*} \frac{\norm{\mathbf{\Phi}_{\G^{\sharp c}}^\ast
        \mathbf{r}}^2_{\ell^\infty(\R^{\G^{\sharp c}})}}
        {\norm{\mathbf{\Phi}^\ast_{\G^\sharp} \mathbf{r}}^2_{\ell^\infty(\R^{\G^\sharp})}}  &
= &   \frac{\norm{\mathbf{\Phi}_\la^\ast
            \mathbf{r}}^2_{\ell^\infty(\R^\la)}}{\norm{\mathbf{\Phi}^\ast_{\G^\sharp}
                \mathbf{r}}^2_{\ell^\infty(\R^{\G^\sharp})}}
            \leq \frac{\norm{\mathbf{\Phi}_{\la}^\ast \mathbf{r}}^2_{\ell^2(\R^\la)}}{\frac{1}{k}\norm{\mathbf{\Phi}^\ast_{\G^\sharp} \mathbf{r}}^2_{\ell^2(\R^{\G^\sharp})}} \\
& = & \frac{k\norm{\mathbf{\Phi}_{\la}^\ast \mathbf{\Phi}_{\G^\sharp} \mathbf{y}_{\G^\sharp}}^2_{\ell^2(\R^\la)}}
                {\norm{\mathbf{\Phi}^\ast_{\G^\sharp} \mathbf{\Phi}_{\G^\sharp} \mathbf{y}_{\G^\sharp}}^2_{\ell^2(\R^{\G^\sharp})}}
       \leq \frac{k\de_{k+1}^2}{(1-\de_k)^2}\,
\end{eqnarray*}
where the first inequality is due to usual norm inequalities and the
second by (\ref{e:Almost_orth}) in the numerator and consecutive
application of (\ref{e:PHIstar_below}) and the left-hand side of
(\ref{e:RIP_deltaK}) in the denominator. Comparing the right side of
the last inequality with $\al^2$ ends the proof.
\hfill $\blacksquare$ \vskip .5cm   

\begin{cor}\label{c:ERC_Weak_PHIstar_below_Cor1} With the
conditions of Theorem \ref{t:ERC_Weak_PHIstar_below}, if any of the
algorithms has selected $k$ elements, then it has found the whole
support of the $k$-sparse vector $\mathbf{x}$.
\end{cor}

Since SWOMP, as well as OMP, select new elements at each iteration, in at most $k$ steps
they recover the $k$-sparse signal by \eqref{reconstruction}. Thus we have

\begin{cor}\label{c:ERC_Weak_PHIstar_below_Cor2}
With the conditions of Theorem \ref{t:ERC_Weak_PHIstar_below}, the
SWOMP and OMP algorithms recover every $k$-sparse vector
$\mathbf{x}$ in at most $k$ iterations.
\end{cor}

\begin{rem}\label{r:I_limitations}
Observe that $\delta_k\leq\delta_{k+1}$ since the set of all
$k$-sparse vectors is contained in the set of all $(k+1)$-sparse
vectors. Moreover, to achieve (\ref{e:cond_selecGP}) the RIP
constants $\delta_{k+1}$ and $\delta_k$ must satisfy
$\delta_{k+1}<\alpha(1-\delta_k)/\sqrt{k}$. This gives a restriction
on $k$, that is $k<(\alpha(1-\delta_k)/\delta_{k+1})^2$.
\end{rem}

\begin{rem}\label{r:I_vs_ERC_incoherent}
The proof of Theorem \ref{t:ERC_Weak_PHIstar_below} follow the ideas
of the proof in \cite{Tr1}, regarding deterministic
\emph{quasi-incoherent dictionaries}, which uses the fact that
$\mathbf{r}=P_{\G^\sharp}
\mathbf{r}=\mathbf{\Phi}_{\G^\sharp}\mathbf{\Phi}_{\G^\sharp}^\dagger
\mathbf{r}$ (that is OMP type algorithms) and then bounds with usual norm inequalities. The
condition obtained is then called \emph{Exact Reconstruction
Condition} $ERC_\al$ for \emph{quasi-incoherent dictionaries} since
the result is on OMP and therefore exact reconstruction is
guaranteed in at most $k$ iterations; a similar result is called
\emph{Stability Condition} in \cite{GrVa} since it is applied to MP
and no exact reconstruction of sparse signals is guaranteed in $k$
iterations in this case. In our case, a straight use of the
consequences of the RIP property lead us to the result.
\end{rem}

\begin{rem}\label{r:I_vs_ERC_incoherent}
Suppose that $\delta_{k+1} < \frac{1}{1+\sqrt{k}}\,.$ Since $\delta_k \leq \delta_{k+1}$ we have
$$
\delta_{k+1} < \frac{1}{1+\sqrt{k}} = \frac{1 - \frac{1}{1+\sqrt{k}} }{\sqrt{k}} \leq \frac{1 - \delta_k}{\sqrt{k}}\,.
$$
Thus, condition \eqref{e:cond_selecGP} is satisfied with $\alpha =
1\,.$ By Corollary \ref{c:ERC_Weak_PHIstar_below_Cor2}, if
$\delta_{k+1} < \frac{1}{1+\sqrt{k}}\,$ the OMP algorithm recovers
any $k$-sparse vector $\bf x$ in at most $k$ iterations. This result
has recently appeared in \cite{Maleh} and \cite{MS} (see also
\cite{DW10} and \cite{LT} for previous smaller bounds on
$\delta_{k+1}$). Moreover, it is proved in \cite{MS} that there
exits a $k$-sparse vector $\bf x$ $\in \mathbb R^{k+1}$ and a
$(k+1)\times (k+1)$ matrix $\Phi$ satisfying RIP with
$\delta_{k+1}=\frac{1}{\sqrt{k}}$ such that OMP does not recover
$\bf x$ in $k$ iterations (proving a conjecture estated in
\cite{DM09}). Thus the bound $\delta_{k+1} < \frac{1}{1+\sqrt{k}}$
is nearly optimal.
\end{rem}

\

\section{Support Identification with a Relaxed Weak Selection
Rule}\label{Sec:Select_on_Rlx_Weak_GPMPOMP}
Next, we consider
another decision rule to select indices in the true support of a
sparse signal. With the same notation as in section \ref{Sec:SupId_RIP} , let
\begin{equation}\label{e:Sel_Rule_tilde}
    \mathcal{\tilde{I}}_n := \mathcal{\tilde{I}}_n (\tilde\alpha):=\left\{ i: \abs{g^n_i} \ge \tilde\al \norm{\mathbf{r}^{n-1}}_{\ell^2(\R^m)}\right\},
\end{equation}
be the \textbf{\emph{relaxed weak}} selection rule. Rule
(\ref{e:Sel_Rule_tilde}) compares the absolute value of
$g_i^n=\ip{\phi_i}{\mathbf{r}^{n-1}}$ with
$\norm{\mathbf{r}^{n-1}}_{\ell^2(\mathbb{R}^m)}$. It can be proved
(see the proof of Theorem 9.10 in \cite{Mal}, p. 422) that there
exists $\beta_0>0$ such that for any $\mathbf{x}\in\mathbb{R}^m$,
$\sup_{i=1,\ldots,N}\abs{\ip{\phi_i}{\mathbf{x}}}\geq\beta_0\norm{\mathbf{x}}_{\ell^2(\mathbb{R}^m)}$.
Thus, $\tilde{\mathcal{I}}_n(\al)$ is a non empty set if we choose
$\alpha\leq\beta_0$.

\

Selection rule (\ref{e:Sel_Rule_tilde}) is similar to the ones considered in  \cite{DTDS} and \cite{CoDaDe2}.
In  \cite{DTDS} the rule is
$$\mathcal{J}_n (t):=\left\{ i: \abs{g^n_i} \ge \frac{t}{\sqrt{m}} \norm{\mathbf{r}^{n-1}}_{\ell^2(\R^m)}\right\},$$
and the algorithm developed is called StOMP. Thus, $\mathcal{\tilde{I}}_n (\tilde\alpha)$ is $\mathcal{J}_n (t)$ with
$\tilde\alpha = t/\sqrt{m}$. In \cite{CoDaDe2} the rule is
$$\mathcal{D}_n (\delta):=\left\{ i: \abs{g^n_i} \ge \frac{t}{\sqrt{k}} \norm{\mathbf{r}^{n-1}}_{\ell^2(\R^m)}\right\},$$
and the algorithm developed is called DTresh (and its cousin STresh).
Thus, $\mathcal{\tilde{I}}_n (\tilde\alpha)$ is $\mathcal{D}_n (t)$ with
$\tilde\alpha = t/\sqrt{k}$. In both of these algorithms the update of the approximation is done using the same updating as the OMP algorithm.

\

In rule \eqref{e:Sel_Rule_tilde}  we compare with the energy of the residual
$\norm{\mathbf{r}^{n-1}}_{\ell^2(\R^m)}$ instead of comparing with
the maximum of the correlations of the residue with the ``atoms" of
$\mathbf{\Phi}$, \emph{i.e.} $\norm{\mathbf{\Phi}^\ast
\mathbf{r}^{n-1}}_{\ell^\infty(\R^N)}$. We choose the term
``\textbf{relaxed}" since, as we will see in Theorem
\ref{t:ERCNew_Weak_PHIstar_below}, the condition on $\delta_{k+1}$
and $\delta_k$ is weaker than the one required in Theorem
\ref{t:ERC_Weak_PHIstar_below}. 
Therefore, we decided to
call the greedy algorithms with this selection rule
\textbf{\emph{Relaxed}} WGP, WMP or WOMP algorithms, writing RWGP,
RWMP and RWOMP respetively.

\

We still consider $\mathbf{x}$ a $k-$sparse vector with support on $\G^\sharp$.

\begin{thm}\label{t:ERCNew_Weak_PHIstar_below}
Let $\mathbf{\Phi}$ satisfies RIP with $\de_{k+1}$. A sufficient
condition for the \emph{RWGP, RWMP} and \emph{RWOMP} algorithms to
identify elements on the support $\G^\sharp$ of $\mathbf{x}$ with
the \textbf{\emph{relaxed}} weak selection rule $\mathcal{\tilde{I}}_n (\tilde\alpha)$ given by
(\ref{e:Sel_Rule_tilde}) is that $\mathcal{\tilde{I}}_n (\tilde\alpha) \neq \emptyset$ and
\begin{equation}\label{e:cond_Rlx_selecGP}
\tilde\al > \frac{\de_{k+1}}{(1-\de_k)^{1/2}}.
\end{equation}
\end{thm}
\textbf{\emph{Proof.}} It is basically the same proof as for Theorem
\ref{t:ERC_Weak_PHIstar_below}. Again, we drop the superindex of
$\mathbf{r}$ and make the assumption $\mathbf{r}\in\text{span
}(\mathbf{\Phi}_{\G^\sharp})$ to proceed by induction. The condition $\mathcal{\tilde{I}}_n
\subset \Gamma^\sharp$ is implied by
$$\norm{\mathbf{\Phi}_{\G^{\sharp c}}^\ast \mathbf{r}}_{\ell^\infty(\G^{\sharp c})} < \tilde\al \norm{\mathbf{r}}_{\ell^2(\R^m)}.$$
As before, rearranging to express it as a quotient (called
\emph{relaxed weak greedy selection ratio}), squaring and choosing
$\la$ as one index in $\G^{\sharp c}$ with the largest value of
$\mathbf{\Phi}_{\G^{\sharp c}}^\ast \mathbf{r}$, it gives
\begin{eqnarray*}
\frac{\norm{\mathbf{\Phi}_{\G^{\sharp c}}^\ast \mathbf{r}}^2_{\ell^\infty(\R^{\G^{\sharp
c}})}}{\norm{\mathbf{r}}^2_{\ell^2(\R^m)}}
&= &   \frac{\norm{\mathbf{\Phi}_\la^\ast \mathbf{r}}^2_{\ell^\infty(\R^\la)}}{\norm{\mathbf{\Phi}_{\G^\sharp} \mathbf{y}_{\G^\sharp}}^2_{\ell^2(\R^m)}}
        = \frac{\norm{\mathbf{\Phi}_{\la}^\ast \mathbf{\Phi}_{\G^\sharp} \mathbf{y}_{\G^\sharp}}^2_{\ell^2(\R^\la)}}
            {\norm{\mathbf{\Phi}_{\G^\sharp} \mathbf{y}_{\G^\sharp}}^2_{\ell^2(\R^m)}} \\
& \leq & \frac{\norm{\mathbf{\Phi}_{\la}^\ast \mathbf{\Phi}_{\G^\sharp}}^2_{\ell^2(\R^{\G^\sharp})\to \ell^2(\R^\la)}
                \norm{\mathbf{y}_{\G^\sharp}}^2_{\ell^2(\R^{\G^\sharp})}}
            {\norm{\mathbf{\Phi}_{\G^\sharp} \mathbf{y}_{\G^\sharp}}^2_{\ell^2(\R^m)}} \\
& \leq &   \frac{\de_{k+1}^2}{(1-\de_k)},
\end{eqnarray*}
where the first and second inequalities are due to usual norm
inequalities and the third by (\ref{e:Almost_orth}) in the numerator
and the left side of (\ref{e:RIP_deltaK}) in the denominator.
Comparing the left hand side of the last inequality with $\tilde\al^2$
ends the proof.
\hfill $\blacksquare$ \vskip .5cm   


\begin{rem}\label{New1}
Suppose $\mathbf{r} \in span \,  ({\mathbf{\Phi}_{\G^{\sharp}}})$ where $\mathbf{\Phi}$ is a RIP matrix;
then
$$ \norm{\mathbf{\Phi}_{\G^{\sharp }}^\ast \mathbf{r}}_{\ell^\infty(\R^{\G^{\sharp}})}
\geq \frac{1}{\sqrt{k}} \norm{\mathbf{\Phi}_{\G^{\sharp }}^\ast \mathbf{r}}^2_{\ell^2(\R^{\G^{\sharp}})}
\geq \frac{(1-\delta_k)^{1/2}}{\sqrt{k}} \norm{ \mathbf{r}}^2_{\ell^2(\R^{\G^{\sharp}})},
$$
by the usual norm inequality and Lemma \ref{l:PHIstar_below}. Thus taking
\begin{equation} \label{Neq2}
\tilde \al \leq \frac{(1-\delta_k)^{1/2}}{\sqrt{k}}
\end{equation}
we always have  $\mathcal{\tilde{I}}_n (\tilde\alpha) \neq \emptyset $ at each iteration of the algorithms with
the Relaxed selection rule. Notice that \eqref{e:cond_Rlx_selecGP} and \eqref{Neq2} could hold at the same time
for some value of $\tilde \al$ only if
$$ \frac{\delta_{k+1}}{(1-\delta_k)^{1/2}} \leq \frac{(1-\delta_k)^{1/2}}{\sqrt{k}}, $$
which gives the following restriction: $\sqrt{k} \leq \frac{1-\delta_k}{\delta_{k+1}}.$
\end{rem}

\begin{rem}\label{r:I_tlida_no_restrctn}
Theorem \ref{t:ERCNew_Weak_PHIstar_below} could be considered as an
Exact Reconstruction Condition for the \textbf{\emph{Relaxed}} Weak
OMP (RWOMP) algorithm with RIP matrices and with parameter $\de_{k+1}$ (see
Corollary \ref{c:ERC_Weak_PHIstar_below_Cor2}) as long as \eqref{Neq2} is satisfied. The probability of
success depends exclusively on the probability that a random
ensemble verifies RIP. It has been proved that Gaussian, Bernoulli
and partial Fourier matrices verify RIP with very high probability
(exponential concentration) as long as the number of measurements
$m\geq Ck\log(N/k)$ for the first two ensembles (see
\cite{CanTao2},\cite{CanTao1}) and $m\geq C k\log^5(N)$ for the
random Fourier (see \cite{CanTao2}, \cite{RuVe1}).
\end{rem}



\section{Convergence. The Sparse Case.}\label{Sec:Convrgnc}
In this section we obtain convergence rates for the
 SWGP, SWMP and SWOMP algorithms and their \textbf{relaxed}
counterparts for matrices satisfying RIP. The results here are given in terms of the reduccion of the energy
of the residuals  $\mathbf{r}^n = \mathbf{y} - \mathbf{y^n}$ of the observation $\mathbf{y}=\mathbf{\Phi} \mathbf{x}$
rather in the energy of  the residuals of the approximation, $\mathbf{x} - \mathbf{x}^n$, as is done in \cite{NeTr} for CoSaMP
and in \cite{CoDaDe2} for DThresh.

\subsection{Convergence of GP, SWGP and RWGP}\label{Sec:GP_conv} In
\cite{BlDa1} the analysis of convergence of the GP is based on an
existence theorem (see Theorem 9.10 in \cite{Mal}) when
$\mathbf{\Phi}\in\C^{N\times N_1}$ is a dictionary (this means that
$\mathbf{\Phi}$ contains at least a base for $\R^N$ and thus
$N_1\geq N$) that verifies the \emph{Exact Reconstruction Condition}
ERC$_\al(\G)$ for \emph{quasi-incoherent} dictionaries (see
\cite{Tr1}). The analysis in \cite{TrGi} is done for random
\emph{admissible} matrices. The result in the theorem below is
obtained for matrices satisfying RIP.
We still consider $\mathbf{x}$ a $k-$sparse vector with support on $\G^\sharp$.

\begin{thm}\label{t:Convrg_GP_PHIstar_below}
Consider the algorithms \emph{GP, SWGP} and \emph{RWGP} and suppose
that at iteration $n$ we have $\G^s\subset\G^\sharp$,
$s=1,2,\ldots,n$. Let $\mathbf{\Phi}$ verifies RIP with $\de_k$.
Then, for all $k$-sparse vectors $\mathbf{x}\in\mathbb{R}^N$ (supp
$(\mathbf{x})=\Gamma^\sharp$),
\begin{equation}\label{e:Convrg_GP_PHIstar_below}
\norm{\mathbf{r}^n}^2_{\ell^2(\R^m)} \le C_k
\norm{\mathbf{r}^{n-1}}^2_{\ell^2(\R^m)},
\end{equation}
with $C_k= (1- \frac{1-\de_k}{k(1+\de_k)})^{1/2} <1$. In the case
RWGP suppose $\tilde{\alpha}\leq\frac{(1-\delta_k)^{1/2}}{\sqrt{k}}$
so that $\tilde{\mathcal{I}}_n(\tilde{\alpha})\neq\emptyset$.
\end{thm}
\emph{\textbf{Proof.}} To shorten notation we will write
$\mathbf{d}^n=\mathbf{d}^n_{\Gamma^n}$. We have
\begin{eqnarray}\label{e:Convrg_GP_PHIstar_below_0}
\nonumber
    & &\!\!\!\!\!\!\!\!\!\!\norm{\mathbf{r}^n}^2_{\ell^2(\R^m)}\\
\nonumber
    &=&    \ip{\mathbf{r}^{n-1}-a^n\mathbf{\Phi}_{\Gamma^n}\mathbf{d}^n}{\mathbf{r}^{n-1}-a^n\mathbf{\Phi}_{\Gamma^n}\mathbf{d}^n} \\
\nonumber
    & = & \norm{\mathbf{r}^{n-1}}^2_{\ell^2(\R^m)}
            - {a^n}\ip{\mathbf{r}^{n-1}}{\mathbf{\Phi}_{\Gamma^n}\mathbf{d}^n}
            - a^n\ip{\mathbf{\Phi}_{\G^n}\mathbf{d}^n}{\mathbf{r}^{n-1}}
            + \ip{a^n\mathbf{\Phi}_{\Gamma^n}\mathbf{d}^n}{a^n\mathbf{\Phi}_{\Gamma^n}\mathbf{d}^n}\\
\nonumber
    & = &   \norm{\mathbf{r}^{n-1}}^2_{\ell^2(\R^m)} - 2\frac{\abs{\ip{\mathbf{\Phi}_{\G^n} \mathbf{d}^n_{\G^n}}{\mathbf{r}^{n-1}}}^2}
                    {\norm{\mathbf{\Phi}_{\Gamma^n}\mathbf{d}^n}^2_{\ell^2(\R^m)}}
                    + \frac{\abs{\ip{\mathbf{\Phi}_{\Gamma^n}\mathbf{d}^n}{\mathbf{r}^{n-1}}}^2}{\norm{\mathbf{\Phi}_{\Gamma^n}\mathbf{d}^n}^4_{\ell^2(\R^m)}}
        \norm{\mathbf{\Phi}_{\Gamma^n}\mathbf{d}^n}^2_{\ell^2(\R^m)}\\
    & = & \norm{\mathbf{r}^{n-1}}^2_{\ell^2(\R^m)}
        -\frac{\abs{\ip{\mathbf{r}^{n-1}}{\mathbf{\Phi}_{\Gamma^n}\mathbf{d}^n}}^2}{\norm{\mathbf{\Phi}_{\Gamma^n}\mathbf{d}^n}^2_{\ell^2(\R^m)}}.
\end{eqnarray}
Since $\mathbf{d}^n=\mathbf{\Phi}^\ast_{\Gamma^n}\mathbf{r}^{n-1}$,
the second term above can be bounded below by
\begin{eqnarray}\label{e:Convrg_GP_PHIstar_below_1}
\nonumber
\frac{\abs{\ip{\mathbf{r}^{n-1}}{\mathbf{\Phi}_{\Gamma^n}\mathbf{d}^n}}^2}{\norm{\mathbf{\Phi}_{\Gamma^n}\mathbf{d}^n}^2_{\ell^2(\R^m)}}
    & = &   \frac{\abs{\ip{\mathbf{\Phi}_{\G^n}^\ast \mathbf{r}^{n-1}}{\mathbf{d}^n}}^2}{\norm{\mathbf{\Phi}_{\G^n}
            \mathbf{d}^n}^2_{\ell^2(\R^m)}} \\
\nonumber
    & \ge &   \frac{\norm{\mathbf{\Phi}^\ast_{\Gamma^n}\mathbf{r}^{n-1}}^4_{\ell^2(\R^{\G^n})}}
        {\norm{\mathbf{\Phi}_{\Gamma^n}}^2_{\ell^2(\R^{\G^n})\to\ell^2(\R^m)} \norm{\mathbf{\Phi}^\ast_{\Gamma^n} \mathbf{r}^{n-1}}^2_{\ell^2(\R^{\G^n})}}\\
    & \ge & \frac{\norm{\mathbf{\Phi}^\ast_{\Gamma^n}\mathbf{r}^{n-1}}^2_{\ell^2(\R^{\Gamma^n})}}{1+\de_k}.
\end{eqnarray}
We have
\begin{equation}\label{e:Convrg_GP_PHIstar_below_2}
    \norm{\mathbf{\Phi}^\ast_{\Gamma^n}\mathbf{r}^{n-1}}_{\ell^\infty(\mathbb{R}^{\Gamma^n})}
    \geq\norm{\mathbf{\Phi}^\ast_{\mathcal{I}_n}\mathbf{r}^{n-1}}_{\ell^\infty(\mathbb{R}^{\mathcal{I}_n})}
    =\norm{\mathbf{\Phi}^\ast\mathbf{r}^{n-1}}_{\ell^\infty(\mathbb{R}^N)},
\end{equation}
because $\Gamma^n\supset\mathcal{I}_n$ in the inequality and because
the definition of $\mathcal{I}_n$ (in GP and SWGP) in the equality.
Since we are supposing that $\Gamma^n\subset\Gamma^\sharp$, using
(\ref{e:Convrg_GP_PHIstar_below_2}) yields
\begin{equation}\label{e:Convrg_GP_PHIstar_below_3}
\norm{\mathbf{\Phi}^\ast_{\Gamma^n}\mathbf{r}^{n-1}}_{\ell^2(\mathbb{R}^{\Gamma^n})}
    \geq \norm{\mathbf{\Phi}^\ast_{\Gamma^n}\mathbf{r}^{n-1}}_{\ell^\infty(\mathbb{R}^{\Gamma^n})}
    \geq \norm{\mathbf{\Phi}^\ast \mathbf{r}^{n-1}}_{\ell^\infty(\mathbb{R}^N)}
    \geq \norm{\mathbf{\Phi}^\ast_{\Gamma^\sharp}
    \mathbf{r}^{n-1}}_{\ell^\infty(\mathbb{R}^{\Gamma^\sharp})}.
\end{equation}
By (\ref{e:Convrg_GP_PHIstar_below_3}) and usual norm inequality
$$\norm{\mathbf{\Phi}^\ast_{\Gamma^n}\mathbf{r}^{n-1}}_{\ell^2(\mathbb{R}^{\Gamma^n})}
    \geq \norm{\mathbf{\Phi}^\ast_{\Gamma^\sharp}\mathbf{r}^{n-1}}_{\ell^\infty(\mathbb{R}^{\Gamma^\sharp})}
    \geq \frac{1}{k}\norm{\mathbf{\Phi}^\ast_{\Gamma^\sharp}\mathbf{r}^{n-1}}_{\ell^2(\mathbb{R}^{\Gamma^\sharp})}.$$
Since $\mathbf{r}^n\in\text{span }(\mathbf{\Phi}_{\Gamma^\sharp})$
(assuming conditions in Theorem \ref{t:ERC_Weak_PHIstar_below} are
met and because $\mathbf{r}^{n}=\mathbf{y}-\mathbf{y}^n$ in GP and
SWGP), by Lemma \ref{l:PHIstar_below} we can write
\begin{equation}\label{e:Convrg_GP_PHIstar_below_4}
\norm{\mathbf{\Phi}^\ast_{\Gamma^n}\mathbf{r}^{n-1}}^2_{\ell^2(\mathbb{R}^{\Gamma^n})}
    \geq \frac{1-\delta_k}{k}
    \norm{\mathbf{r}^{n-1}}^2_{\ell^2(\mathbb{R}^m)}.
\end{equation}
Substituting (\ref{e:Convrg_GP_PHIstar_below_4}) in
(\ref{e:Convrg_GP_PHIstar_below_1}) and
(\ref{e:Convrg_GP_PHIstar_below_0}) one gets
$$\norm{\mathbf{r}^n}^2_{\ell^2(\mathbb{R}^m)}
    \leq \left(1-\frac{1-\delta_k}{k(1+\delta_k)}\right)\norm{\mathbf{r}^{n-1}}^2_{\ell^2(\mathbb{R}^m)},$$
which shows the result for GP and SWGP algorithms.

For RWGP the selection rule
$$\tilde{\mathcal{I}}_n(\tilde{\alpha})=\{i:\abs{\ip{\phi_i}{\mathbf{r}^{n-1}}}
    \geq \tilde{\alpha}\norm{\mathbf{r}^{n-1}}_{\ell^2(\mathbb{R}^m)}\}$$
do not allow us to write (\ref{e:Convrg_GP_PHIstar_below_2}). But in
this case using Remark \ref{New1} we can write
\begin{eqnarray*}
  \norm{\mathbf{\Phi}^\ast_{\Gamma^n}\mathbf{r}^{n-1}}^2_{\ell^2(\mathbb{R}^{\Gamma^n})}
    &\geq& \norm{\mathbf{\Phi}^\ast_{\Gamma^n}\mathbf{r}^{n-1}}^2_{\ell^\infty(\mathbb{R}^{\Gamma^n})}
        \geq \norm{\mathbf{\Phi}^\ast_{\tilde{\mathcal{I}}_n}\mathbf{r}^{n-1}}^2_{\ell^\infty(\mathbb{R}^{\tilde{\mathcal{I}}_n})} \\
    &\geq& \frac{1-\delta_k}{k}
        \norm{\mathbf{r}^{n-1}}^2_{\ell^2(\mathbb{R}^m)},
\end{eqnarray*}
which shows (\ref{e:Convrg_GP_PHIstar_below_4}) for RWGP and
therefore the result.
\hfill $\blacksquare$ \vskip .5cm   

\begin{rem}\label{r:Cnvrgc_WGP_CS_VS_WGP_RndMtx}
In \cite[Theorem 3]{BlDa1}, it is proved that
$\norm{\mathbf{r}^n}_{\ell^2(\mathbb{R}^m)}^2\leq
c\norm{\mathbf{r}^{n-1}}_{\ell^2(\mathbb{R}^m)}^2$ with
$c=(1-\frac{\omega}{\norm{\mathbf{\Phi}}^2_2})$ and $\omega$ is a
positive real number such that $\norm{\mathbf{\Phi}
\mathbf{x}}_{\ell^\infty(\mathbb{R}^N)}^2> \omega
\norm{\mathbf{x}}_{\ell^2(\mathbb{R}^m)}^2$ for all
$\mathbf{x}\in\mathbb{R}^m$. Our Theorem
\ref{t:Convrg_GP_PHIstar_below} gives a value of $C_k$ depending on
the restricted isometry constant $\delta_k$ and the sparseness $k$.
This value of $C_k$ is less than $1$, but very close to $1$ when $k$
increases. It is easy to show that it can never hold $C_k\leq1/2$
when $k\geq 2$.
\end{rem}

Using Theorems \ref{t:ERC_Weak_PHIstar_below},
\ref{t:ERCNew_Weak_PHIstar_below} and
\ref{t:Convrg_GP_PHIstar_below} we deduce the following:

\begin{cor}\label{Cor4.3}
Suppose that $\mathbf{\Phi}$ satisfies RIP with constants $\de_k$ and $\de_{k+1}$.

i) Suppose $ \frac{\sqrt{k} \delta_{k+1}}{1-\delta_k} <1$; then, the
GP algorithm satisfies \eqref{e:Convrg_GP_PHIstar_below}.

ii) Let $0 < \alpha \leq 1$ and suppose $ \frac{\sqrt{k}
\delta_{k+1}}{1-\delta_k} <\al$; then, the SWGP algorithm with
selection rule $\mathcal{I}_n (\alpha)$ satisfies
\eqref{e:Convrg_GP_PHIstar_below}.

iii) Suppose that $\frac{\delta_{k+1}}{1-\delta_k} <
\frac{1}{\sqrt{k}}$ and that $\tilde\al$ is chosen such that $
\frac{\delta_{k+1}}{(1-\delta_k)^{1/2}} < \tilde\al <
\frac{(1-\delta_k)^{1/2}}{\sqrt{k}}$; then, the RWGP algorithm with
selection rule $\mathcal{\tilde I}_n (\tilde\alpha)$ satisfies
\eqref{e:Convrg_GP_PHIstar_below}.
\end{cor}

\begin{rem}\label{r:Cnvrgc_WGP_CS_VS_WGP_RndMtx_2}
Convergence in algorithms of type Gradient Pursuit (GP, WGP y RWGP)
is given in Theorem \ref{t:Convrg_GP_PHIstar_below} in terms of the
convergence of the energy of residual
$\mathbf{r}^n=\mathbf{y}-\mathbf{y}^n$. If $\mathbf{\Phi}$ satisfies
RIP, convergence of the residual in terms of estimation
$\mathbf{x}-\mathbf{x}^n$ can be obtained from the last result. For
algorithms of type GP it is not hard to show that
$\mathbf{y}^n=\mathbf{\Phi}\mathbf{x}^n$. Then,
\begin{eqnarray}\label{e:r:Cnvrgc_WGP_CS_VS_WGP_RndMtx_2_1}
\nonumber
  \norm{\mathbf{y}-\mathbf{y}^n}_{\ell^2(\mathbb{R}^m)}
    &=& \norm{\mathbf{\Phi}\mathbf{x}-\mathbf{\Phi}\mathbf{x}^n}_{\ell^2(\mathbb{R}^m)}
        =\norm{\mathbf{\Phi}_{\Gamma^\sharp}(\mathbf{x}-\mathbf{x}^n)}_{\ell^2(\mathbb{R}^m)} \\
    &\geq& (1-\delta_k)^{1/2}\norm{\mathbf{x}-\mathbf{x}^n}_{\ell^2(\mathbb{R}^N)}
\end{eqnarray}
because the left-hand side of RIP as long as
$\Gamma^n\subset\Gamma^\sharp$ (which is verified under the
conditions of Theorems \ref{t:ERC_Weak_PHIstar_below} and
\ref{t:ERCNew_Weak_PHIstar_below}). Analogously, if we also assume
$\Gamma^{n-1}\subset\Gamma^\sharp$ we have
\begin{eqnarray}\label{e:r:Cnvrgc_WGP_CS_VS_WGP_RndMtx_2_2}
\nonumber
  \norm{\mathbf{y}-\mathbf{y}^{n-1}}_{\ell^2(\mathbb{R}^m)}
    &=& \norm{\mathbf{\Phi}\mathbf{x}-\mathbf{\Phi}\mathbf{x}^{n-1}}_{\ell^2(\mathbb{R}^m)}
        =\norm{\mathbf{\Phi}_{\Gamma^\sharp}(\mathbf{x}-\mathbf{x}^{n-1})}_{\ell^2(\mathbb{R}^m)} \\
    &\leq& (1+\delta_k)^{1/2}\norm{\mathbf{x}-\mathbf{x}^{n-1}}_{\ell^2(\mathbb{R}^N)}
\end{eqnarray}
because the right-hand side of RIP. From
(\ref{e:r:Cnvrgc_WGP_CS_VS_WGP_RndMtx_2_1}) and
(\ref{e:r:Cnvrgc_WGP_CS_VS_WGP_RndMtx_2_2}) we deduce
\begin{equation}\label{e:r:Cnvrgc_WGP_CS_VS_WGP_RndMtx_2_3}
\norm{\mathbf{x}-\mathbf{x}^n}_{\ell^2(\mathbb{R}^N)} \leq
    C_k(\frac{1+\delta_k}{1-\delta_k})^{1/2}\norm{\mathbf{x}-\mathbf{x}^{n-1}}_{\ell^2(\mathbb{R}^N)},
\end{equation}
where $C_k$ is the constant in Theorem
\ref{t:Convrg_GP_PHIstar_below}. Observe that, besides conditions of
Corollary \ref{Cor4.3}, if we want to assure convergence in
$\ell^2(\mathbb{R}^N)$ of $\mathbf{x}^n$ to $\mathbf{x}$ we need
$$(\frac{1+\delta_k}{1-\delta_k})^{1/2}(1-\frac{1-\delta_k}{k(1+\delta_k)})^{1/2}<1,$$
which requires $\delta_k<\frac{1}{2k+1}$.
\end{rem}

If from some iteration we had $\Gamma^n=\Gamma^\sharp=\text{supp
}(\mathbf{x})$ reduction of energy of residuals would be faster than
the given in (\ref{e:Convrg_GP_PHIstar_below}), as the next result
shows.

\begin{thm}\label{t:Convrg_GP_PHIstar_below_SppIdentfied}
Consider GP, WGP and RWGP algorithms and suppose that at iteration
$n_0$ we have $\Gamma^{n_0}=\Gamma^\sharp=\text{sop }(\mathbf{x})$.
Suppose that
$\alpha,\tilde{\alpha}>\frac{\delta_{k+1}}{(1-\delta_k)^{1/2}}$ and
$\mathbf{\Phi}$ verifies RIP with parameter $\delta_{k+1}$. Then,
for all $n\geq n_0$,
\begin{equation}\label{e:Convrg_GP_PHIstar_below_SppIdentfied}
\norm{\mathbf{r}^n}_{\ell^2(\mathbb{R}^m)}\leq
D_k\norm{\mathbf{r}^{n-1}}_{\ell^2(\mathbb{R}^m)}
\end{equation}
with
$D_k=(1-\frac{1-\delta_k}{1+\delta_k})^{1/2}=(\frac{2\delta_k}{1+\delta_k})^{1/2}<1.$
\end{thm}
\textbf{Proof}. Equality (\ref{e:Convrg_GP_PHIstar_below_0}) and
inequality (\ref{e:Convrg_GP_PHIstar_below_1}) in the proof of
Theorem \ref{t:Convrg_GP_PHIstar_below} are still valid in our
context. Since $\Gamma^{n_0}=\Gamma^\sharp$ we have
$\Gamma^n=\Gamma^\sharp$ for $n\geq n_0$. The fact that
$\alpha,\tilde{\alpha}>\frac{\delta_{k+1}}{(1-\delta_k)^{1/2}}$
allow us to conclude $\tilde{\mathcal{I}}_{n_0},
\tilde{\mathcal{I}}_{n}\subset\Gamma^\sharp$.

Hence, we can replace $\Gamma^n$ by $\Gamma^\sharp$ in
(\ref{e:Convrg_GP_PHIstar_below_1}) and since
$\mathbf{r}^{n-1}\in\text{span }(\mathbf{\Phi}_{\Gamma^\sharp})$, by
Lemma \ref{l:PHIstar_below} we can write
\begin{eqnarray*}
  \norm{\mathbf{\Phi}^\ast_{\Gamma^n}\mathbf{r}^{n-1}}^2_{\ell^2(\mathbb{R}^m)}
    &=& \norm{\mathbf{\Phi}^\ast_{\Gamma^\sharp}\mathbf{r}^{n-1}}^2_{\ell^2(\mathbb{R}^m)} \\
    &\geq&
    (1-\delta_k)\norm{\mathbf{r}^{n-1}}^2_{\ell^2(\mathbb{R}^m)}.
\end{eqnarray*}
Substituting this inequality in (\ref{e:Convrg_GP_PHIstar_below_0})
and (\ref{e:Convrg_GP_PHIstar_below_1}) we obtain the result for the
three algorithms.
\hfill $\blacksquare$ \vskip .5cm   

\begin{rem}\label{r:t:Convrg_GP_PHIstar_below_SppIdentfied}
Constant $D_k$ of Theorem
\ref{t:Convrg_GP_PHIstar_below_SppIdentfied} can be made as close to
$0$ as desired taking $\delta_k$ small enough. In particular, it is
enough to take $\delta_k\leq 1/7\approx 0.143$ to reduce the energy
of the residuals by half in just one iteration, if conditions of
Theorem \ref{t:Convrg_GP_PHIstar_below} are satisfied.

Reasoning as in Remark \ref{r:Cnvrgc_WGP_CS_VS_WGP_RndMtx_2} we have
$$\norm{\mathbf{x}-\mathbf{x}^n}_{\ell^2(\mathbb{R}^N)}\leq
(\frac{1+\delta_k}{1-\delta_k})^{1/2}(1-\frac{1-\delta_k}{1+\delta_k})^{1/2}
\norm{\mathbf{x}-\mathbf{x}^{n-1}}_{\ell^2(\mathbb{R}^N)}$$ for
which it is enough to take $\delta_k<1/3$ to assure convergence of
$\mathbf{x}^n$ to $\mathbf{x}$ in $\ell^2(\mathbb{R}^N)$ once we
have $\Gamma^{n_0}=\Gamma^\sharp$.
\end{rem}

\

\subsection{Convergence of SWMP and RWMP}\label{Sec:MP_conv}
The results on convergence in this section are similar to those
obtained in \cite{GrVa} for quasi-incoherent dictionaries.

\begin{thm}\label{t:Convrg_WMP_PHIstar_below}
Consider the algorithms SWMP and RWMP. Suppose that conditions of
Theorems \ref{t:ERC_Weak_PHIstar_below} (for SWMP) and
\ref{t:ERCNew_Weak_PHIstar_below} are verified so that
$\Gamma^s\subset\Gamma^\sharp$, $s=1,2,\ldots$. Then, for every
$k$-sparse vector $\mathbf{x}\in\mathbb{R}^N$ with supp
$(\mathbf{x})=\Gamma^\sharp$,
\begin{equation}\label{e:Convrg_WMP_PHIstar_below}
\norm{\mathbf{r}^n}^2_{\ell^2(\R^m)}\leq
C'_k\norm{\mathbf{r}^{n-1}}^2_{\ell^2(\R^m)},
\end{equation}
with $C'_k=(1-\frac{(1-\delta_k)^2}{k})^{1/2}<1.$
\end{thm}
\textbf{Proof.} We have for SWMP as well as for RWMP algorithms that
\begin{eqnarray}\label{e:Convrg_WMP_PHIstar_below_1}
\nonumber
 \norm{\mathbf{r}^n}^2_{\ell^2(\R^m)}
    &=& \ip{\mathbf{r}^{n-1}-\mathbf{\Phi}_{\mathcal{I}_n}\mathbf{\Phi}_{\mathcal{I}_n}^\ast\mathbf{r}^{n-1}}
        {\mathbf{r}^{n-1}-\mathbf{\Phi}_{\mathcal{I}_n}\mathbf{\Phi}_{\mathcal{I}_n}^\ast\mathbf{r}^{n-1}} \\
\nonumber
    &=& \norm{\mathbf{r}^{n-1}}^2_{\ell^2(\R^m)}
        -\ip{\mathbf{r}^{n-1}}{\mathbf{\Phi}_{\mathcal{I}_n}\mathbf{\Phi}_{\mathcal{I}_n}^\ast\mathbf{r}^{n-1}}\\
\nonumber
    & & -\ip{\mathbf{\Phi}_{\mathcal{I}_n}\mathbf{\Phi}_{\mathcal{I}_n}^\ast\mathbf{r}^{n-1}}{\mathbf{r}^{n-1}}
        + \norm{\mathbf{\Phi}_{\mathcal{I}_n}\mathbf{\Phi}_{\mathcal{I}_n}^\ast \mathbf{r}^{n-1}}^2_{\ell^2(\R^m)}\\
    &=& \norm{\mathbf{r}^{n-1}}^2_{\ell^2(\R^m)}
        - 2 \norm{\mathbf{\Phi}_{\mathcal{I}_n}^\ast\mathbf{r}^{n-1}}^2_{\ell^2(\R^{\mathcal{I}_n})}
       + \norm{\mathbf{\Phi}_{\mathcal{I}_n}\mathbf{\Phi}_{\mathcal{I}_n}^\ast
       \mathbf{r}^{n-1}}^2_{\ell^2(\R^m)}.
\end{eqnarray}
By (\ref{e:PHIstar_above}) we have
$\norm{\mathbf{\Phi}_{\mathcal{I}_n}\mathbf{\Phi}_{\mathcal{I}_n}^\ast\mathbf{r}^{n-1}}^2_{\ell^2(\mathbb{R}^m)}
\leq
(1+\delta_k)\norm{\mathbf{\Phi}_{\mathcal{I}_n}^\ast\mathbf{r}^{n-1}}^2_{\ell^2(\mathbb{R}^{\mathcal{I}_n})}$
since $\mathcal{I}_n\subset\Gamma^\sharp$ and
$\abs{\Gamma^\sharp}=k$. Hence, from
(\ref{e:Convrg_WMP_PHIstar_below_1}) we have
\begin{eqnarray}\label{e:Convrg_WMP_PHIstar_below_2}
\nonumber
  \norm{\mathbf{r}^n}^2_{\ell^2(\mathbb{R}^m)}
    &\leq& \norm{\mathbf{r}^{n-1}}^2_{\ell^2(\mathbb{R}^m)} -
        2\norm{\mathbf{\Phi}_{\mathcal{I}_n}^\ast\mathbf{r}^{n-1}}^2_{\ell^2(\mathbb{R}^{\mathcal{I}_n})}
        + (1+\delta_k)
        \norm{\mathbf{\Phi}_{\mathcal{I}_n}^\ast\mathbf{r}^{n-1}}^2_{\ell^2(\mathbb{R}^{\mathcal{I}_n})}\\
    &=& \norm{\mathbf{r}^{n-1}}^2_{\ell^2(\mathbb{R}^m)}
        -(1-\delta_k)\norm{\mathbf{\Phi}_{\mathcal{I}_n}^\ast\mathbf{r}^{n-1}}^2_{\ell^2(\mathbb{R}^{\mathcal{I}_n})}.
\end{eqnarray}
For SWMP as well as RWMP we have
\begin{equation}\label{e:Convrg_WMP_PHIstar_below_3}
\norm{\mathbf{\Phi}_{\mathcal{I}_n}^\ast\mathbf{r}^{n-1}}^2_{\ell^2(\mathbb{R}^{\mathcal{I}_n})}
\geq
\frac{1-\delta_k}{k}\norm{\mathbf{r}^{n-1}}^2_{\ell^2(\mathbb{R}^m)}.
\end{equation}
For SWMP the proof is as in (\ref{e:Convrg_GP_PHIstar_below_4}) from
Theorem \ref{t:Convrg_GP_PHIstar_below} and it is not hard to prove
it also for RWMP.

Substituting (\ref{e:Convrg_WMP_PHIstar_below_3}) in
(\ref{e:Convrg_WMP_PHIstar_below_2}) we have
$$\norm{\mathbf{r}^n}^2_{\ell^2(\mathbb{R}^m)} \leq
(1-\frac{(1-\delta_k)^2}{k})\norm{\mathbf{r}^{n-1}}^2_{\ell^2(\mathbb{R}^m)}$$
which is the desired result.
\hfill $\blacksquare$ \vskip .5cm   

\begin{cor}\label{c:t:Convrg_WMP_PHIstar_below}
Suppose $\mathbf{\Phi}$ satisfies RIP with constants $\delta_k$ and
$\delta_{k+1}$.
\begin{enumerate}
  \item[a)] Let $0<\alpha\leq 1$ and suppose that
  $\frac{\sqrt{k}\delta_{k+1}}{1-\delta_k}<\alpha$; then, SWMP with selection rule $\mathcal{I}_n(\alpha)$
  satisfies (\ref{e:Convrg_WMP_PHIstar_below}).
  \item[b)] Suppose that $\frac{\delta_{k+1}}{1-\delta_k}<\frac{1}{\sqrt{k}}$
  and
  $\tilde{\alpha}$ is chosen so that $$\frac{\delta_{k+1}}{(1-\delta_k)^{1/2}}<\tilde{\alpha}<\frac{(1-\delta_k)^2}{\sqrt{k}};$$
  then, RWMP with selection rule
  $\tilde{\mathcal{I}}_n(\tilde{\alpha})$ satisfies
  (\ref{e:Convrg_WMP_PHIstar_below}).
\end{enumerate}
\end{cor}

\begin{rem}\label{r:t:Convrg_WMP_PHIstar_below_1}
Constant $C'_k$ of Theorem \ref{t:Convrg_WMP_PHIstar_below} is a
numer less than $1$, so there is always a decreasing in the residual
energy. However, it is close to $1$ when $k$ increases. It is easy
to show that we can never have $C'_k\leq 1/2$ when $k\geq 2$.
\end{rem}

\begin{rem}\label{r:t:Convrg_WMP_PHIstar_below_2}
As in Remark \ref{r:Cnvrgc_WGP_CS_VS_WGP_RndMtx_2} the decreasing of
the residual energy given in (\ref{e:Convrg_WMP_PHIstar_below}) for
SWMP y RWMP can be translated to convergence of the estimation.
Thus,
$$\norm{\mathbf{x}-\mathbf{x}^n}_{\ell^2(\mathbb{R}^N)} \leq (\frac{1+\delta_k}{1-\delta_k})^{1/2}
C'_k \norm{\mathbf{x}-\mathbf{x}^{n-1}}_{\ell^2(\mathbb{R}^N)}.$$
For
$(\frac{1+\delta_k}{1-\delta_k})^{1/2}(1-\frac{(1-\delta_k)^2}{k})^{1/2}$
to be less than $1$ we must have
$$1-\frac{(1-\delta_k)^2}{k}<\frac{1-\delta_k}{1+\delta_k}\Leftrightarrow \frac{2\delta_k}{1+\delta_k}<\frac{(1-\delta_k)^2}{k}.$$
Since
$(1+\delta_k)(1-\delta_k)^2=(1-\delta_k)(1-\delta_k^2)=1-\delta_k-\delta_k^2+\delta_k^3$
last inequality is equivalent to
$$2k\delta_k<1-\delta_k-\delta_k^2+\delta_k^3 \Leftrightarrow \delta_k^3-\delta_k^2-(1+2k)\delta_k+1>0.$$
If $\delta_k<\frac{1}{2k+2}$, we have
$$1>\delta_k+(2k+1)\delta_k>\delta_k^2+(2k+1)\delta_k>\delta_k^2+(2k+1)\delta_k-\delta_k^3,$$
from which
$$\delta_k<\frac{1}{2k+2}$$
is enough to have convergence of approximations in WMP y RWMP.
\end{rem}

\subsection{Convergencia de WOMP y RWOMP}\label{sS:CS_conv_WOMP_RWOMP}
In WOMP as well as in RWOMP the residual $\mathbf{r}^n$ is the
vector that carries out the distance from $\mathbf{y}$ to the
subspace
$V_n=\{\mathbf{\Phi}_{\Gamma^n}\mathbf{x}:\mathbf{x}\in\mathbb{R}^n
\text{ with supp}(\mathbf{x})\subset\Gamma^n \}$. Since
$\Gamma^{n-1}\subset\Gamma^n$, it is clear that
$$\norm{\mathbf{r}^n}_{\ell^2(\mathbb{R}^m)}\leq
 \norm{\mathbf{r}^{n-1}}_{\ell^2(\mathbb{R}^m)}$$
is always accomplished. A strict inequality can be obtained
observing that the residual in WOMP or RWOMP, temporarily denoted
$\mathbf{r}^n_{OMP}$, has an energy no larger than that for WGP,
RWGP, WMP or RWMP, temporarily denoted $\mathbf{r}^n_{GP}$ and
$\mathbf{r}^n_{MP}$, since $\mathbf{y}^n_{GP}$ as well as
$\mathbf{y}^n_{MP}$ are elements from $V_n$.

Therefore, if conditions of Theorems \ref{t:ERC_Weak_PHIstar_below}
are verified (for WGP) or \ref{t:ERCNew_Weak_PHIstar_below} (for
RWGP) we can apply Theorem \ref{t:Convrg_GP_PHIstar_below} to get
\begin{eqnarray}\label{e:conv_r_OMP}
\nonumber
  \norm{\mathbf{r}^n_{OMP}}_{\ell^2(\mathbb{R}^m)}
    &\leq& \norm{\mathbf{r}^n_{GP}}_{\ell^2(\mathbb{R}^m)} \leq C_k\norm{\mathbf{r}^{n-1}_{GP}}_{\ell^2(\mathbb{R}^m)}\\
    &\leq& C_k^n\norm{\mathbf{r}^0_{GP}}_{\ell^2(\mathbb{R}^m)}
        \leq C_k^n\norm{\mathbf{y}}_{\ell^2(\mathbb{R}^m)},
\end{eqnarray}
with $C_k<1$, the constant of Theorem
\ref{t:Convrg_GP_PHIstar_below}.

Analogously, but using Theorem \ref{t:Convrg_WMP_PHIstar_below} we
have
\begin{equation}\label{e:conv_r_OMP_2}
    \norm{\mathbf{r}^n_{OMP}}_{\ell^2(\mathbb{R}^m)}
    \leq C_k^{'n} \norm{\mathbf{y}}_{\ell^2(\mathbb{R}^m)},
\end{equation}
with $C'_k<1$ the constant of Theorem
\ref{t:Convrg_WMP_PHIstar_below}. Among the constants $C_k$ and
$C_k'$ the relation is $C_k<C'_k$ since
\begin{eqnarray*}
  C_k<C'_k
    &\Leftrightarrow& \frac{1-\delta_k}{k(1+\delta_k)}>\frac{(1-\delta_k)^2}{k}
        \Leftrightarrow \frac{1}{1+\delta_k}>1-\delta_k \\
    &\Leftrightarrow& 1>1-\delta_k^2,
\end{eqnarray*}
and the last inequality is true because $0<\delta_k<1$. Therefore,
(\ref{e:conv_r_OMP}) gives faster convergence than
(\ref{e:conv_r_OMP_2}) and it proves that GP algorithms converge
faster than MP algorithms.

Since $\mathbf{r}^n_{OMP}$ is a vector perpendicular to $V_n$, the
constant in (\ref{e:conv_r_OMP}) could be improved, at least in
principle. Since
$$\mathbf{r}^n_{OMP}=\mathbf{y}-\mathbf{y}^n=\mathbf{y}-\mathbf{\Phi}_{\Gamma^n}\mathbf{\Phi}_{\Gamma^n}^\dag\mathbf{y},$$
where $\mathbf{\Phi}_{\Gamma^n}^\dag$ is the pseudo-inverse of
$\mathbf{\Phi}_{\Gamma^n}$, we have to study
\begin{eqnarray}\label{e:conv_r_OMP_3}
\nonumber
  \norm{\mathbf{r}^n_{OMP}}^2_{\ell^2(\mathbb{R}^m)}
    &=& \norm{\mathbf{y}}^2_{\ell^2(\mathbb{R}^m)}
        -\ip{\mathbf{y}}{\mathbf{\Phi}_{\Gamma^n}\mathbf{\Phi}_{\Gamma^n}^\dag\mathbf{y}}
        -\ip{\mathbf{\Phi}_{\Gamma^n}\mathbf{\Phi}_{\Gamma^n}^\dag\mathbf{y}}{\mathbf{y}}\\
    &+&\norm{\mathbf{\Phi}_{\Gamma^n}\mathbf{\Phi}_{\Gamma^n}^\dag\mathbf{y}}^2_{\ell^2(\mathbb{R}^m)}.
\end{eqnarray}

We are not yet able to find a bound for (\ref{e:conv_r_OMP_3}) of
the form
\begin{equation}\label{e:conv_r_OMP_4}
\norm{\mathbf{r}^n_{OMP}}_{\ell^2(\mathbb{R}^m)}\leq
B_k^n\norm{\mathbf{y}}_{\ell^2(\mathbb{R}^m)}
\end{equation}
with $B_K<C_k$. In the case that conditions of Theorems
\ref{t:ERC_Weak_PHIstar_below} and \ref{t:ERCNew_Weak_PHIstar_below}
are satisfied (for example, if
$\delta_{k+1}<\alpha\frac{1-\delta_k}{\sqrt{k}}$ for WOMP and
$\frac{\delta_{k+1}}{(1-\delta_k)^{1/2}}<\tilde{\alpha}<\frac{(1-\delta_k)^{1/2}}{\sqrt{k}}$
for RWOMP) inequalities (\ref{e:conv_r_OMP}) and
(\ref{e:conv_r_OMP_2}) are trivial if $n\geq k$ since algorithms of
type OMP identify the support of a $k$-sparse signal $\mathbf{x}$ in
as much $k$ iterations, and then $\mathbf{r}^n=0$.

Therefore, it is only interesting to find bounds of the form
(\ref{e:conv_r_OMP_4}) from expression (\ref{e:conv_r_OMP_3}) if it
is satisfied with values of $\delta_{k+1}$ and $\delta_k$ less
restrictive than those in Theorems \ref{t:ERC_Weak_PHIstar_below}
and \ref{t:ERCNew_Weak_PHIstar_below}, for which algorithms do not
identify the support of $\mathbf{x}$.


\vskip0.7cm \section{Behavior of the selection rules for some random
matrices}\label{sS:CS_RndMtx} The reader can find in \cite{DeV2} a
way to construct matrices that satisfy RIP deterministically. These
matrices are of order $m\times N$ with $m=p^2$ ($p$ a prime number)
and $N=p^{r+1}$, $0<r<p$, and satisfy RIP with $k<\frac{p}{r}+1$ and
$\delta_k=(k-1)r/p$. Therefore, $m=p^2>(k-1)^2r^2>(k-1)^2$; which
gives a value of $m$ much larger than the necessary to recover a
$k$-sparse signal with the $\ell^1$ minimization which is $m\geq
Ck\log(N/k)$.

It is known (see \cite{BDDW}) that there exist random matrices that
satisfy RIP with parameter $\delta_k$ for any $m\geq Ck\log(N/k)$
with probability greater than $1-2e^{-c'm}$. Among those are the
matrices that satisfy an inequality known as \textbf{concentration
of measure}, namely,
\begin{equation}\label{e:Conc_of_Measure_RndMtx}
\Prb{\abs{\norm{\mathbf{\Phi}(\omega)\mathbf{x}}_{\ell^2(\mathbb{R}^m)}-\norm{\mathbf{x}}_{\ell^2(\mathbb{R}^N)}}\geq
\varepsilon\norm{\mathbf{x}}^2_{\ell^2(\mathbb{R}^N)}}\leq
2e^{mc_0(\varepsilon)}, \;\;\; 0<\varepsilon<1,
\end{equation}
where the probability is taken over all random matrices
$\mathbf{\Phi}(\omega)$ of order $m\times N$ and
$c_0(\varepsilon)>0$ is a constant that depends only on
$\varepsilon$.

An example of such matrices are those $\mathbf{\Phi}=(\phi_{i,j})$
such that $\phi_{i,j}$ is an independent Gaussian random variable
$N(0,1/\sqrt{m})$, that is, with mean $0$ and standard deviation
$1/\sqrt{m}$. In this case we have
$c_0(\varepsilon)=\frac{\varepsilon^2}{4}-\frac{\varepsilon^3}{6}$
(see \cite{DasGup}).

Another example are the matrices whose entries are independent
Bernoulli random variables with values $\{-1/\sqrt{m},1/\sqrt{m}\}$,
with probability $1/2$ each. In this case we also have
$c_0(\varepsilon)=\frac{\varepsilon^2}{4}-\frac{\varepsilon^3}{6}$
(see \cite{Ach}).

In this section we will study the behavior of the selection rules
and algorithms given in \ref{Sec:grdy_algthms} and
\ref{Sec:Select_on_Rlx_Weak_GPMPOMP} with respect to the random
matrices $\mathbf{\Phi}$ as just described above. The aim is to
prove directly that this kind of matrices select elements of the
support of a $k$-sparse signal with high probability, following the
reasoning given in \cite{TrGi} for \emph{Orhogonal Matching Pursuit}
(OMP).

We start with a result on random processes (see \cite{TrGi} and the
references cited in this article), whose proof is given for
completeness:
\begin{lem}\label{l:Bnd_IP_Rnd_Procss}
\begin{enumerate}
  \item[a)] Let $\mathbf{z}$ be a vector with dimension $m$ whose components are Gaussian r.v. $N(0,1/\sqrt{m})$ i.i.d.
    Independently a vector $\mathbf{u}$ unitary in $\ell^2(\mathbb{R}^m)$ is chosen. We have, for $0<\varepsilon\leq 1$,
    \begin{equation}\label{e:Bnd_IP_Rnd_Procss_gauss}
        \Prb{\abs{\ip{\mathbf{u}}{\mathbf{z}}}\geq\varepsilon}\leq e^{-\frac{\varepsilon^2}{2}m}.
    \end{equation}
  \item[b)] Let $\mathbf{w}$ a vector of dimension $m$ whose entries are symmetric Bernoulli r.v.
    $\{-1/\sqrt{m},1/\sqrt{m}\}$ i.i.d.
    Independently a vector $\mathbf{u}$ unitary in
    $\ell^2(\mathbb{R}^m)$ is chosen. We have, for $0<\varepsilon\leq 1$,
    \begin{equation}\label{e:Bnd_IP_Rnd_Procss_bern}\end{equation}
    $$\Prb{\abs{\ip{\mathbf{u}}{\mathbf{w}}}\geq\varepsilon}\leq 2e^{-\frac{\varepsilon^2}{2}m}.$$
\end{enumerate}
\end{lem}
\textbf{Proof}. The inner product
$\ip{\mathbf{u}}{\mathbf{z}}=\sum_{i=1}^mu_iz_i$ is a Gaussian r.v.
(the sum of Gaussian r.v. is a Gaussian r.v.) with mean
$\E{\ip{\mathbf{u}}{\mathbf{z}}}=\sum_{i=1}^m u_i\E{z_i}=0$ and
standard deviation
$$(\E{\abs{\ip{\mathbf{u}}{\mathbf{z}}}^2})^{1/2}
    =(\sum_{i=1}^m u_i^2 \E{z_i^2} +\sum_{i=1}^m\sum_{j\neq i} u_iu_j\E{z_iz_j})^{1/2}=\frac{1}{\sqrt{m}},$$
since the r.v. $z_i$ are independents and that $\mathbf{u}$,
unitary, is independent from $\mathbf{z}$. Hence, since
$\abs{\ip{\mathbf{u}}{\mathbf{z}}}$ is symmetric
\begin{equation}\label{e:Bnd_IP_Rnd_Procss_gauss_proof}
    \Prb{\abs{\ip{\mathbf{u}}{\mathbf{z}}}\geq \varepsilon}
        =2\Prb{\ip{\mathbf{u}}{\mathbf{z}}\geq\varepsilon}=2\frac{\sqrt{m}}{\sqrt{2\pi}}
            \int_\varepsilon^\infty e^{-\frac{1}{2}mx^2}dx
        =\sqrt{\frac{2}{\pi}}\int_{\varepsilon\sqrt{m}}^\infty e^{-\frac{y^2}{2}} dy
\end{equation}
making the change of variable $\sqrt{m}x=y$. Let
$$I=\int_{\varepsilon\sqrt{m}}^\infty e^{-\frac{y^2}{2}} dy.$$
We have
\begin{eqnarray*}
  I^2
    &=& \left(\int_{\varepsilon\sqrt{m}}^\infty e^{-\frac{y^2}{2}} dy\right)\left(\int_{\varepsilon\sqrt{m}}^\infty e^{-\frac{x^2}{2}} dx\right)
        =\int_{\varepsilon\sqrt{m}}^\infty\left(\int_{\varepsilon\sqrt{m}}^\infty e^{-\frac{y^2+x^2}{2}} dy\right)dx \\
    &\leq& \int_{R_{\varepsilon,m}}\int e^{-\frac{y^2+x^2}{2}} dx dy,
\end{eqnarray*}
where $$R_{\varepsilon,m}=\{(x,y)=w\in\mathbb{R}^2:x,y\geq
0,\norm{w}_2\geq\varepsilon\sqrt{2m}\}.$$ Passing to polar
coordinates
$$I^2\leq\int_0^{\pi/2}\int_{\varepsilon\sqrt{2m}}^\infty e^{-\frac{r^2}{2}}r dr d\theta
    = \frac{\pi}{2}\left[-e^{-\frac{r^2}{2}}\right]_{\varepsilon\sqrt{2m}}^\infty
    = \frac{\pi}{2} e^{-\frac{\varepsilon^2m}{2}}.$$
Therefore, $I\leq
\sqrt{\frac{\pi}{2}}e^{-\frac{\varepsilon^2m}{2}}$, which
substituting in (\ref{e:Bnd_IP_Rnd_Procss_gauss_proof}) yields
(\ref{e:Bnd_IP_Rnd_Procss_gauss}).

b) In this case the Hoeffding inequality can be applied (see Theorem
4 in \cite{Lu}) to get
\begin{eqnarray*}
  \Prb{\abs{\ip{\mathbf{u}}{\mathbf{w}}\geq\varepsilon}}
    &=& 2\Prb{\sum_{i=1}^m u_iw_i \geq\varepsilon}
        = 2\Prb{\sum_{i=1}^m u_iw_i -\E{\sum_{i=1}^m u_iw_i}\geq \varepsilon} \\
    &=& 2e^{-2\varepsilon^2/\sum_{i=1}^m (u_i/\sqrt{m}- (-u_i/\sqrt{m}))^2}
        =2e^{-\frac{\varepsilon^2 m}{2}},
\end{eqnarray*}
since $\E{\sum_{i=1}^m u_iw_i}=\sum_{i=1}^m u_i\E{w_i}=0$ because
$\mathbf{u}$ is independent from the r.v. $w_i$ and
$\norm{\mathbf{u}}_2=1$.
\hfill $\blacksquare$ \vskip .5cm   

In this section we will consider random matrices
$\mathbf{\Phi}(\omega)\in\mathbb{R}^{m\times N}$ that satisfy next
conditions, similar to those conditions for admissible matrices in
\cite{TrGi}:
\begin{enumerate}
  \item[(M1)] The columns of $\mathbf{\Phi}(\omega)$ are
    statistically independents.
  \item[(M2)] For each column $\phi_j(\omega)$, $j=1,\ldots,N$, of
    $\mathbf{\Phi}(\omega)$ we have
    $\E{\norm{\phi_j(\omega)}^2_{\ell^2(\mathbb{R}^m)}}=1$.
  \item[(M3)] Let $\mathbf{u}\in\mathbb{R}^m$ a vector with $\norm{\mathbf{u}}_{\ell^2(\mathbb{R}^m)}\leq
    1$. If $\phi(\omega)$ is a column of
    $\mathbf{\Phi}(\omega)$ independent from $\mathbf{u}$,
    $$\Prb{\abs{\ip{\phi(\omega)}{\mathbf{u}}}\geq \varepsilon}\leq q_1e^{-c_1\varepsilon^2 m},$$
    with $q_1,c_1$ constants, $q_1\geq 1$.
  \item[(M4)] For every set $\Gamma\subset\{1,\ldots,N\}$ with
    $\abs{\Gamma}\leq k<N$ and for every $\mathbf{r}\in\text{span }(\mathbf{\Phi}_\Gamma(\omega))$
    we have
    $$\Prb{\norm{\mathbf{\Phi}^\ast_\Gamma(\omega)\mathbf{r}}_{\ell^2(\mathbb{R}^\Gamma)}\geq \frac{1}{2}\norm{r}_{\ell^2(\mathbb{R}^m)}}
        \geq 1-q_2D^k e^{-c_2 m},$$
    with $q_2,D$ and $c_2$ constants, $q_2,D>1$.
\end{enumerate}

These properties are satisfied by the Gaussian and Bernoulli random
matrices  as described at the beginning of this section. Property
(M3) es the content of Lemma \ref{l:Bnd_IP_Rnd_Procss} (observe that
if $\norm{\mathbf{u}}_2\leq 1$ Lemma \ref{l:Bnd_IP_Rnd_Procss} is
also verified since if we substitute $\mathbf{u}$ for a unitary
vector in its direction then the probability increases) and
condition (M4) is proved in Lemma 5.1 in \cite{BDDW}, where a proof
is given from a concentration of measure inequality as that in
(\ref{e:Conc_of_Measure_RndMtx}).

\subsection{Probabilistic support identification for relaxed algorithms}\label{sS:CS_RndMtx_Id_RlxAlgs}
In this section we show that matrices satisfying (M1), (M2), (M3)
and (M4) allow to identify indices in the support of a $k$-sparse
signal with high probability. The result follows the arguments in
\cite{TrGi}, with necessary modifications to fit the relaxed
selection rule $\tilde{I}(\tilde{\alpha})$ given in
(\ref{e:Sel_Rule_tilde}).

\begin{thm}\label{t:RndMtx_RlxWkSelRul}
Let $\mathbf{\Phi}(\omega)\in \mathbb{R}^{m\times N}$ be a random
matrix satisfyiing (M1), (M2), (M3) and (M4). Let
$\mathbf{x}\in\mathbb{R}^N$ with $\text{supp
}(\mathbf{x})=\Gamma^\sharp$ y $\abs{\Gamma^\sharp}\leq k<N$. Sea
$\mathbf{y}=\mathbf{\Phi}\mathbf{x}$. Suppose that
$\tilde{\alpha}\leq 1/2\sqrt{k}$ and given $l$, $1\leq l<N$,
\begin{equation}\label{e:RndMtx_RlxWkSelRul_1}
    m\geq \max\{\frac{1}{c_1\tilde{\alpha}^2}\ln q_3l(N-k), \frac{2k}{c_2}\ln
    D\},
\end{equation}
with $q_3=q_1+q_2$. Algorithms RWMP, RWOMP y RWGP, with selection
rule $\tilde{I}_n(\tilde{\alpha})$ given by(\ref{e:Sel_Rule_tilde}),
identify elements from $\Gamma^\sharp$ in the first $l$ iterations
with probability greater or equal to
$$1-q_3 l(N-k)e^{-c_1 \tilde{\alpha}^2m}.$$
\end{thm}
\textbf{Proof}. For any of the given algorithms with selection rule
$\tilde{I}_n(\tilde{\alpha})$, let $E_n$ be the set of matrices that
identify elements from $\Gamma^\sharp=\text{sop }(\mathbf{x})$ at
iteration $n$, $n=1,2,3,\ldots$.

We start bounding $\Prb{E_1}$, that is, the probability that a
matrix identify elements from $\Gamma^\sharp$ in the first
iteration. For a random matrix $\mathbf{\Phi}(\omega)$ to belong to
$E_1$ it is enough that the next two conditions are satisfied:
\begin{enumerate}
  \item[$(A_1)$] That $\tilde{I}_1(\tilde{\alpha})\not=\emptyset$.
  \item[$(B_1)$] For
    $\mathbf{u}_0=\mathbf{r}^0/\norm{\mathbf{r}^0}_{\ell^2(\mathbb{R}^m)}$
    (we remind that $\mathbf{r}^0=\mathbf{y}$) it should be true
    that
    $$\frac{\norm{\mathbf{\Phi}^\ast_{\Gamma^{\sharp c}}(\omega)\mathbf{r}^0}_{\ell^\infty(\mathbb{R}^{\Gamma^{\sharp c}})}}
        {\norm{\mathbf{r}^0}_{\ell^2(\mathbb{R}^m)}}
        =\norm{\mathbf{\Phi}^\ast_{\Gamma^{\sharp c}}(\omega)\mathbf{u}_0}_{\ell^\infty(\mathbb{R}^{\Gamma^{\sharp c}})}
        =\max_{j\in\Gamma^{\sharp
        c}}\abs{\ip{\mathbf{u}_0}{\phi_j(\omega)}}<\tilde{\alpha}.$$
\end{enumerate}
With abuse of notation, we call $A_1$ and $B_1$ the sets of matrices
that satisfy $(A_1)$ and $(B_1)$ respectively. We have
\begin{equation}\label{e:RndMtx_RlxWkSelRul_2}
\Prb{E_1}\geq \Prb{A_1\cap B_1}=\Prb{A_1|B_1}\Prb{B_1}.
\end{equation}
To estimate $\Prb{B_1}$ we write
\begin{eqnarray*}
  \Prb{B_1}
    &=& \Prb{\max_{j\in\Gamma^{\sharp c}}\abs{\ip{\mathbf{u}_0}{\phi_j(\omega)}}<\tilde{\alpha}}
        =\Prb{\cap_{j\in\Gamma^{\sharp c}}\{\abs{\ip{\mathbf{u}_0}{\phi_j(\omega)}}<\tilde{\alpha}\}}\\
    &=& \prod_{j\in\Gamma^{\sharp c}} \Prb{\abs{\ip{\mathbf{u}_0}{\phi_j(\omega)}}<\tilde{\alpha}}
\end{eqnarray*}
due to the independence of the columns $\phi_j$ from
$\mathbf{\Phi}(\omega)$ expressed in (M1). Since
$\mathbf{u}_0=\mathbf{y}/\norm{\mathbf{y}}_2$ is unitary and
$\mathbf{y}=\mathbf{\Phi}\mathbf{x}=\mathbf{\Phi}_{\Gamma^\sharp}\mathbf{x}_{\Gamma^\sharp}$
only depends on the columns $\phi_j$ with $j\in\Gamma^\sharp$,
$\mathbf{u}_0$ is independent from the columns $\phi_j$ with
$j\in\Gamma^{\sharp c}$, and we can use (M3) to get
\begin{equation}\label{e:RndMtx_RlxWkSelRul_3}
\Prb{B_1}\geq \prod_{j\in\Gamma^{\sharp c}}
(1-q_1e^{-c_1\tilde{\alpha}^2 m})=(1-q_1e^{-c_1\tilde{\alpha}^2
m})^{N-k}\geq 1-q_1(N-k)e^{-c_1\tilde{\alpha}^2 m},
\end{equation}
since $(1-x)^n\geq1-nx$ if $n\geq 1$ and $x\leq 1$.

On the other hand, conditioned to $B_1$ we have that
$\mathbf{\Phi}\in A_1\Leftrightarrow
\norm{\mathbf{\Phi}_{\Gamma^\sharp}\mathbf{y}}_\infty\geq\tilde{\alpha}\norm{\mathbf{y}}_2$.
Since $\tilde{\alpha}\leq \frac{1}{2\sqrt{k}}$,
\begin{eqnarray*}
  \Prb{A_1|B_1}
    &=& \Prb{\norm{\mathbf{\Phi}^\ast_{\Gamma^\sharp}(\omega)\mathbf{y}}_{\ell^\infty(\mathbb{R}^{\Gamma^\sharp})}
        \geq \tilde{\alpha}\norm{\mathbf{y}}_{\ell^2(\mathbb{R}^m)}}\\
    &\geq& \Prb{\norm{\mathbf{\Phi}^\ast_{\Gamma^\sharp}(\omega)\mathbf{y}}_{\ell^2(\mathbb{R}^{\Gamma^\sharp})}
        \geq \tilde{\alpha}\sqrt{k}\norm{\mathbf{y}}_{\ell^2(\mathbb{R}^m)}} \\
    &\geq& \Prb{\norm{\mathbf{\Phi}^\ast_{\Gamma^\sharp}(\omega)\mathbf{y}}_{\ell^2(\mathbb{R}^{\Gamma^\sharp})}
        \geq \frac{1}{2}\norm{\mathbf{y}}_{\ell^2(\mathbb{R}^m)}}.
\end{eqnarray*}
Using now (M4) we deduce (observe that
$\mathbf{y}=\mathbf{\Phi}\mathbf{x}$, therefore
$\mathbf{y}\in\text{span }(\mathbf{\Phi}_{\Gamma^\sharp})$):
\begin{equation}\label{e:RndMtx_RlxWkSelRul_4}
\Prb{A_1|B_1}\geq 1-q_2D^ke^{-c_2 m}\geq 1-q_2e^{-c_2 m/2}
\end{equation}
taking $-c_2 m+k\ln D\leq -c_2 m/2$, that is,
$m\geq\frac{2k}{c_2}\ln D$.

Substituting (\ref{e:RndMtx_RlxWkSelRul_3}) and
(\ref{e:RndMtx_RlxWkSelRul_4}) in (\ref{e:RndMtx_RlxWkSelRul_2}) we
get
\begin{eqnarray}\label{e:RndMtx_RlxWkSelRul_5}
\nonumber
  \Prb{E_1}
    &\geq& (1-q_1(N_k)e^{-c_1\tilde{\alpha}^2 m})(1-q_2 e^{-c_2 m/2})\\
\nonumber
    &\geq& 1-q_1(N-k)e^{-c_1\tilde{\alpha}^2 m} - q_2e^{-c_2 m/2} \\
    &\geq& 1-q_3(N-k)e^{-c_1\tilde{\alpha}^2 m},
\end{eqnarray}
since $\tilde{\alpha}<1/2$, with $q_3=q_1+q_2$.

We have proved the result for $l=1$. Suppose the result is true
until iteration $l-1$, this is,
\begin{equation}\label{e:RndMtx_RlxWkSelRul_6}
\Prb{\cap_{n=1}^{l-1}E_n}\geq 1-q_3(l-1)(N-k)e^{-c_1\tilde{\alpha}^2
m}.
\end{equation}
Until iteration $l$ we have
\begin{equation}\label{e:RndMtx_RlxWkSelRul_7}
\Prb{\cap_{n=1}^l E_n}=
\Prb{E_l|\cap_{n=1}^{l-1}E_n}\Prb{\cap_{n=1}^{l-1}E_n}.
\end{equation}
For a random matrix $\mathbf{\Phi}(\omega)$ to belong to $E_l$ it is
enough that next two conditions are satisfied:
\begin{enumerate}
  \item[$(A_l)$] That $\tilde{I}_l(\tilde{\alpha})\not=\emptyset$.
  \item[$(B_l)$] For
        $\mathbf{u}_{l-1}=\mathbf{r}^{l-1}/\norm{\mathbf{r}^{l-1}}_2$
        it should be true that
            $$\frac{\norm{\mathbf{\Phi}^\ast_{\Gamma^{\sharp c}}(\omega)\mathbf{r}^{l-1}}_{\ell^\infty(\mathbb{R}^{\Gamma^{\sharp c}})}}
        {\norm{\mathbf{r}^{l-1}}_{\ell^2(\mathbb{R}^m)}}
        =\norm{\mathbf{\Phi}^\ast_{\Gamma^{\sharp c}}(\omega)\mathbf{u}_{l-1}}_{\ell^\infty(\mathbb{R}^{\Gamma^{\sharp c}})}
        =\max_{j\in\Gamma^{\sharp
        c}}\abs{\ip{\mathbf{u}_{l-1}}{\phi_j(\omega)}}<\tilde{\alpha}.$$
\end{enumerate}
With abuse of notation, we call $A_l$ and $B_l$ the sets of matrices
that satisfy $(A_l)$ and $(B_l)$ respectively. We have
\begin{equation}\label{e:RndMtx_RlxWkSelRul_8}
\Prb{E_l|\cap_{n=1}^{l-1}E_n}\geq
\Prb{A_l|B_l\cap(\cap_{n=1}^{l-1}E_n)}\Prb{B_l|\cap_{n=1}^{l-1}E_n}.
\end{equation}
With the same reasoning that leads to (\ref{e:RndMtx_RlxWkSelRul_3})
we get
\begin{equation}\label{e:RndMtx_RlxWkSelRul_9}
\Prb{B_l|\cap_{n=1}^{l-1}E_n}\geq 1-q_1(N-k)e^{-c_1\tilde{\alpha}^2
m}
\end{equation}
since we are conditioned to the fact that the algorithm has selected
indices from $\Gamma^\sharp$ until iteration $l-1$, we have that
$\mathbf{r}^{l-1}$ only depends on the columns $\phi_j$ with
$j\in\Gamma^\sharp$, therefore $\mathbf{u}_{l-1}$ is a unitary
vector independent from the columns $\phi_j$ with
$j\in\Gamma^{\sharp c}$.

Conditioned to $B_l$, $\mathbf{\Phi}\in A_l\Leftrightarrow
\norm{\mathbf{\Phi}_{\Gamma^\sharp}(\omega)\mathbf{r}^{l-1}}_\infty\geq\tilde{\alpha}\norm{\mathbf{r}^{l-1}}_2$.
Since $\tilde{\alpha}\leq \frac{1}{2\sqrt{k}}$, we have
\begin{eqnarray*}
  \Prb{A_l|B_l\cap(\cap_{n=1}^{l-1}E_n)}
    &=& \Prb{\{\norm{\mathbf{\Phi}^\ast_{\Gamma^\sharp}(\omega)\mathbf{r}^{l-1}}_{\ell^\infty(\mathbb{R}^{\Gamma^\sharp})}
        \geq \tilde{\alpha}\norm{\mathbf{r}^{l-1}}_{\ell^2(\mathbb{R}^m)} \} | \cap_{n=1}^{l-1}E_n }\\
    &\geq& \Prb{\{\norm{\mathbf{\Phi}^\ast_{\Gamma^\sharp}(\omega)\mathbf{y}}_{\ell^2(\mathbb{R}^{\Gamma^\sharp})}
        \geq \tilde{\alpha}\sqrt{k}\norm{\mathbf{r}^{l-1}}_{\ell^2(\mathbb{R}^m)}\} | \cap_{n=1}^{l-1}E_n} \\
    &\geq& \Prb{\norm{\mathbf{\Phi}^\ast_{\Gamma^\sharp}(\omega)\mathbf{r}^{l-1}}_{\ell^2(\mathbb{R}^{\Gamma^\sharp})}
        \geq \frac{1}{2}\norm{\mathbf{r}^{l-1}}_{\ell^2(\mathbb{R}^m)}| \cap_{n=1}^{l-1}E_n}.
\end{eqnarray*}
It is now possible to use (M4) since $\mathbf{r}^{l-1}\in\text{span
}(\mathbf{\Phi}_{\Gamma^\sharp})$ (which is not hard to prove) and
since we are computing the probability conditioned to the fact that
the algorithm has selected indices from $\Gamma^\sharp$. Hence, with
the same reasoning that leads to (\ref{e:RndMtx_RlxWkSelRul_4}) we
get
\begin{equation}\label{e:RndMtx_RlxWkSelRul_10}
\Prb{A_l|B_l\cap(\cap_{n=1}^{l-1}E_n)}\geq 1-q_2e^{-c_2 m/2}.
\end{equation}
From (\ref{e:RndMtx_RlxWkSelRul_9}) and
(\ref{e:RndMtx_RlxWkSelRul_10}) we deduce
\begin{equation}\label{e:RndMtx_RlxWkSelRul_11}
\Prb{E_l|\cap_{n=1}^{l-1}E_n}\geq 1-q_3(N-k)e^{-c_1\tilde{\alpha}^2
m}
\end{equation}
reasoning as in the chain of inequalities that leads to
(\ref{e:RndMtx_RlxWkSelRul_5}). Substitute
(\ref{e:RndMtx_RlxWkSelRul_11}) and (\ref{e:RndMtx_RlxWkSelRul_6})
in (\ref{e:RndMtx_RlxWkSelRul_7}) to get
\begin{eqnarray}
\nonumber
  \Prb{\cap_{n=1}^l E_n}
    &\geq& (1-q_3(N-k)e^{-c_1\tilde{\alpha}^2 m/2})(1-q_3(l-1)(N-k)e^{-c_1\tilde{\alpha}^2 m}) \\
\nonumber
    &\geq& 1-q_3(N-k)e^{-c_1\tilde{\alpha}^2m/2} -q_3(l-1)(N-k)e^{-c_1\tilde{\alpha}^2 m}\\
    &=& 1-q_3l(N-k)e^{-c_1\tilde{\alpha}^2 m},
\end{eqnarray}
which is the desired estimation. Condition
(\ref{e:RndMtx_RlxWkSelRul_1}) assures that this probability is
greater than zero.
\hfill $\blacksquare$ \vskip .5cm   

\begin{rem}\label{r:t:RndMtx_RlxWkSelRul_1}
    Taking $\tilde{\alpha}=\frac{1}{2\sqrt{k}}$, condition
    (\ref{e:RndMtx_RlxWkSelRul_1}) is written
    $$m\geq \max\{\frac{4k}{c_1} \ln q_3l(N-k),\frac{2k}{c_2}\ln D\}.$$
    Hence, if we take $m\geq C k\ln l(N-k)$, with $C$
    large enough, inequality (\ref{e:RndMtx_RlxWkSelRul_1}) is guaranteed.
\end{rem}

For RWOMP with parameter $\tilde{\alpha}$ each iteration adds one
element at least, as long as
$\tilde{\mathcal{I}}_n(\tilde{\alpha})\not=\emptyset$. In this
situation, RWOMP identifies every index within $\Gamma^\sharp$ in as
much $k$ iteration. Besides, the OMP type algorithms recover any
$k$-sparse vector $\mathbf{x}$ once the support is known. Since
$k(N-k)\leq N^2/4$ if $k<N$, we have the following corollary.

\begin{cor}\label{c:t:RndMtx_RlxWkSelRul}
Suppose the same hypothesis that in Theorem
\ref{t:RndMtx_RlxWkSelRul} substituting
(\ref{e:RndMtx_RlxWkSelRul_1}) by
\begin{equation}\label{e:c:t:RndMtx_RlxWkSelRul}
m\geq \max\{\frac{1}{c_1\tilde{\alpha}^2}\ln
q_3\frac{N^2}{4},\frac{2k}{c_2}\ln D\}.
\end{equation}
Then, RWOMP algorithm recovers the $k$-sparse vector $\mathbf{x}$ in
the first $k$ iterations with probability greater or equal to
$$1-q_3\frac{N^2}{4}e^{-c_1\tilde{\alpha}^2 m}.$$
\end{cor}

\begin{rem}\label{r:t:RndMtx_RlxWkSelRul_2}
Suppose we want to obtain the result in Theorem
\ref{t:RndMtx_RlxWkSelRul} with a probability greater than or equal
to $1-\beta$ for a number $\beta\in(0,1)$. It would be enough to
take
$$1-q_3 l(N-k)e^{-c_1\tilde{\alpha}^2 m}\geq 1-\beta.$$
This is accomplished if
$$\beta e^{c_1\tilde{\alpha}^2 m}\geq q_3 l(N-k),$$
for which it is enough to take
\begin{equation}\label{e:r:t:RndMtx_RlxWkSelRul_2}
m\geq \max\{\frac{2}{c_1\tilde{\alpha}^2}\ln \frac{q_3
l(N-k)}{\beta}, \frac{2k}{c_2}\ln D\}.
\end{equation}
To be sure that
$$\ln \frac{q_3 l(N-k)}{\beta}>1$$
for all $q_3\geq 1$, $l<N$ and $k<N$, it suffices to take
$\beta\in(0,1/e)$.

An analogous comment can be done for Corollary
\ref{c:t:RndMtx_RlxWkSelRul}, and in this case
(\ref{e:c:t:RndMtx_RlxWkSelRul}) should be substituted by
\begin{equation}\label{e:r:t:RndMtx_RlxWkSelRul_2}
m\geq \max\{\frac{1}{c_1\tilde{\alpha}^2}\ln \frac{q_3 N^2}{4\beta},
\frac{2k}{c_2}\ln D\}
\end{equation}
for RWOMP to recover a $k$-sparse signal $\mathbf{x}$ within the
first $k$ iterations with probability greater than or equal to
$1-\beta$.
\end{rem}

\subsection{Probabilistic support identification with selection rule
$\mathcal{I}(\alpha)$.}\label{sS:CS_RndMtx_Id_WkAlgs} In this
section we will study the behavior of matrices satisfying (M1),
(M2), (M3) and (M4) with respect to the algorithms with selection
rule
\begin{equation}\label{e:Sel_Rule_RndMtx}
\mathcal{I}_n(\alpha)=\{i:\abs{\ip{\phi_i}{\mathbf{r}^{n-1}}}\geq\alpha\norm{\mathbf{\Phi}^\ast\mathbf{r}^{n-1}}_{\ell^\infty(\mathbb{R}^N)}\},
\end{equation}
$0<\alpha\leq 1$. An advantage of $\mathcal{I}_n(\alpha)$ with
respect to the selection rule
$\tilde{\mathcal{I}}_n(\tilde{\alpha})$ given in
(\ref{e:Sel_Rule_tilde}) is that
$\mathcal{I}_n(\alpha)\not=\emptyset$ for all $\alpha\in(0,1)$ and
for all $n=1,2,\ldots$.

There is a disadvantage, however. If we want to prove that
$\mathcal{I}_n(\alpha)\subset\Gamma^\sharp=\text{supp }(\mathbf{x})$
we need
$$\frac{\norm{\mathbf{\Phi}^\ast_{\Gamma^{\sharp c}}\mathbf{r}^{n-1}}_\infty}
    {\norm{\mathbf{\Phi}^\ast \mathbf{r}^{n-1}}_\infty}<\alpha$$
to be satisfied. If we wrote
$\mathbf{u}_{n-1}=\mathbf{r}^{n-1}/\norm{\mathbf{\Phi}^\ast\mathbf{r}^{n-1}}_\infty$
as in the proof of Theorem \ref{t:RndMtx_RlxWkSelRul} we would not
have $\norm{\mathbf{u}_{n-1}}_2\leq 1$, and could not use (M3).
Property (M4) saves this situation as shown in the next result.

\begin{thm}\label{t:RndMtx_WkSelRul}
Choose $\mathbf{\Phi}(\omega)\in\mathbb{R}^{m\times N}$ a random
matrix satisfying (M1), (M2), (M3) y (M4). Let
$\mathbf{x}\in\mathbb{R}^N$ with supp $(\mathbf{x})=\Gamma^\sharp$
and $\abs{\Gamma^\sharp}\leq k<N$. Let
$\mathbf{y}=\mathbf{\Phi}\mathbf{x}$. Suppose that
\begin{equation}\label{e:RndMtx_WkSelRul}
m\geq \max\{\frac{4k}{c_1\alpha^2}\ln q_3 l(N-k), \frac{2k}{c_2}\ln
D\},
\end{equation}
with $q_3=q_1+q_2$. Algorithms WMP, WOMP and WGP, with selection
rule $\mathcal{I}_n(\alpha)$ given in \ref{e:Sel_Rule_RndMtx},
identify elements from $\Gamma^\sharp$ in the first $l$ iterations
with probability greater than or equal to
$$1-q_3 l(N-k)e^{-c_1\frac{\alpha^2}{4k}m}.$$
\end{thm}
\textbf{Proof}. For any of the given algorithms with selection rule
$\mathcal{I}_n(\alpha)$, let $E_n$ be the set of matrices that
identify elements from $\Gamma^\sharp=\text{supp }(\mathbf{x})$ at
iteration $n$, $n=1,2,\ldots$.

We start bounding $\Prb{E_1}$, that is, the probability that a
matrix identify elements from $\Gamma^\sharp$ in the first
iteration. For a random matrix $\mathbf{\Phi}(\omega)$ belong to
$E_1$ it is enough that
\begin{equation}\label{e:RndMtx_WkSelRul_1}
\frac{\norm{\mathbf{\Phi}^\ast_{\Gamma^{\sharp
c}}(\omega)\mathbf{r}^0}_{\ell^\infty(\mathbb{R}^{\Gamma^{\sharp
c}})}}{\norm{\mathbf{\Phi}^\ast_{\Gamma^\sharp}(\omega)\mathbf{r}^0}_{\ell^\infty(\mathbb{R}^N)}}<\alpha,
\;\;\; (\mathbf{r}^0=\mathbf{y}),
\end{equation}
is verified. Let
$\mathbf{u}_0=\mathbf{y}/2\norm{\mathbf{\Phi}^\ast_{\Gamma^\sharp}\mathbf{y}}$;
we have
\begin{eqnarray}\label{e:RndMtx_WkSelRul_2}
\nonumber
   \frac{\norm{\mathbf{\Phi}^\ast_{\Gamma^{\sharp c}}(\omega)\mathbf{y}}_\infty}
        {\norm{\mathbf{\Phi}^\ast_{\Gamma^\sharp}(\omega)\mathbf{y}}_\infty}
    &\leq& \frac{\sqrt{k}\norm{\mathbf{\Phi}^\ast_{\Gamma^{\sharp c}}(\omega)\mathbf{y}}_\infty}
        {\norm{\mathbf{\Phi}^\ast_{\Gamma^\sharp}(\omega)\mathbf{y}}_2} = 2\sqrt{k}\norm{\mathbf{\Phi}^\ast_{\Gamma^{\sharp c}}(\omega)\mathbf{u}_0}_\infty \\
    &=& 2\sqrt{k}\sup_{j\in\Gamma^{\sharp
        c}}\abs{\ip{\phi_j(\omega)}{\mathbf{u}_0}}.
\end{eqnarray}
Let $A_1$ be the set of matrices satisfying (M1), (M2), (M3) and
(M4) such that
\begin{equation}\label{e:RndMtx_WkSelRul_3}
\sup_{j\in\Gamma^{\sharp
        c}}\abs{\ip{\phi_j(\omega)}{\mathbf{u}_0}}<\frac{\alpha}{2\sqrt{k}}.
\end{equation}
From (\ref{e:RndMtx_WkSelRul_2}) we deduce that if
$\mathbf{\Phi}(\omega)\in A_1$, then $\mathbf{\Phi}(\omega)$ satisfy
(\ref{e:RndMtx_WkSelRul_1}) and we have $\mathbf{\Phi}(\omega)\in
E_1$. Therefore,
\begin{equation}\label{e:RndMtx_WkSelRul_4}
\Prb{E_1}\geq\Prb{A_1}.
\end{equation}
Let $B_1$ be the set of matrices $\mathbf{\Phi}(\omega)$ satisfying
(M1), (M2), (M3) and (M4) such that
\begin{equation}\label{e:RndMtx_WkSelRul_5}
\norm{\mathbf{\Phi}^\ast_{\Gamma^\sharp}(\omega)\mathbf{y}}_2\geq
\frac{1}{2}\norm{\mathbf{y}}_2.
\end{equation}
From (\ref{e:RndMtx_WkSelRul_4}) we deduce
\begin{equation}\label{e:RndMtx_WkSelRul_6}
\Prb{E_1}\geq \Prb{A_1}\geq \Prb{A_1\cap
B_1}=\Prb{A_1|B_1}\Prb{B_1}.
\end{equation}
Since
$\mathbf{y}=\mathbf{\Phi}\mathbf{x}=\mathbf{\Phi}_{\Gamma^\sharp}\mathbf{x}_{\Gamma^\sharp}\in\text{span
}(\mathbf{\Phi}_{\Gamma^\sharp})$ from (M4) we deduce
\begin{equation}\label{e:RndMtx_WkSelRul_7}
\Prb{B_1}\geq 1-q_2 D^k e^{-c_2 m}\geq 1-q_2 e^{-c_2m/2}
\end{equation}
taking $-c_2 m+k\ln D\leq -c_2 m/2$, this is, $m\geq
\frac{2k}{c_2}\ln D$. Conditioned to $B_1$ the vector
$\mathbf{u}_0=\mathbf{y}/2\norm{\mathbf{\Phi}^\ast_{\Gamma^\sharp}\mathbf{y}}_2$
satisfy
$$\norm{\mathbf{u}_0}_2=\frac{1}{2}\frac{\norm{\mathbf{y}}_2}{\norm{\mathbf{\Phi}^\ast_{\Gamma^\sharp}\mathbf{y}}_2}\leq 1.$$
Therefore, to bound
$$\Prb{A_1|B_1}=\Prb{\sup_{j\in\Gamma^{\sharp c}} \abs{\ip{\phi_j(\omega)}{\mathbf{u}_0}}<\frac{\alpha}{2\sqrt{k}}|B_1}$$
we use (M3) to get
\begin{eqnarray*}
  \Prb{A_1|B_1}
    &=& \Prb{\cap_{j\in\Gamma^{\sharp c}} \{\abs{\ip{\phi_j(\omega)}{\mathbf{u}_0}}<\frac{\alpha}{2\sqrt{k}}\} | B_1} \\
    &=& \prod_{j\in\Gamma^{\sharp c}} \Prb{ \abs{\ip{\phi_j(\omega)}{\mathbf{u}_0}}<\frac{\alpha}{2\sqrt{k}} | B_1} \\
    &\geq& (1-q_1 e^{-c_1\frac{\alpha^2}{4k}m})^{N-k}
\end{eqnarray*}
due to the independence of the columns $\phi_j$ of
$\mathbf{\Phi}(\omega)$ expressed in (M1) and because
$\mathbf{u}_0=\mathbf{y}/2\norm{\mathbf{\Phi}^\ast_{\Gamma^\sharp}\mathbf{y}}_2$
only depends on the columns $\phi_j$ with $j\in\Gamma^\sharp$, and
therefore independent of the columns $\phi_j$ with
$j\in\Gamma^{\sharp c}$.

Since $(1-x)^n\geq 1-nx$ if $n\geq 1$, $x\leq 1$, we can write
\begin{equation}\label{e:RndMtx_WkSelRul_8}
\Prb{A_1|B_1}\geq 1-q_1(N-k)e^{-c_1\frac{\alpha^2}{4k}m}.
\end{equation}
Substituting (\ref{e:RndMtx_WkSelRul_8}) and
(\ref{e:RndMtx_WkSelRul_7}) in (\ref{e:RndMtx_WkSelRul_6}) we get
\begin{eqnarray}\label{e:RndMtx_WkSelRul_9}
\nonumber
  \Prb{E_1}
    &\geq& (1-q_1(N-k)e^{-c_1\frac{\alpha^2}{4k}m})(1-q_2e^{-c_1\frac{m}{2}})  \\
\nonumber
    &\geq& 1-q_1(N-k)e^{-c_1\frac{\alpha^2}{4k}m}-q_2e^{-c_1\frac{m}{2}} \\
    &\geq& 1-q_3(N-k)e^{-c_1\frac{\alpha^2}{4k}m}
\end{eqnarray}
with $q_3=q_1+q_2$, since $\frac{\alpha^2}{4k}\leq 1/4<1/2$. This
proves the result for $l=1$.

Suppose that the result is true until iteration $l-1$, this is
\begin{equation}\label{e:RndMtx_WkSelRul_10}
\Prb{\cap_{n=1}^{l-1}E_n} \geq
1-q_3(l-1)(N-k)e^{-c_1\frac{\alpha^2}{4k}m}.
\end{equation}
Until iteration $l$ we have
\begin{equation}\label{e:RndMtx_WkSelRul_11}
\Prb{\cap_{n=1}^{l}E_n} =
\Prb{E_l|\cap_{n=1}^{l-1}E_n}\Prb{\cap_{n=1}^{l-1}E_n}.
\end{equation}
Write
$\mathbf{u}_{l-1}=\mathbf{r}^{l-1}/2\norm{\mathbf{\Phi}^\ast_{\Gamma^\sharp}\mathbf{r}^{l-1}}_2$.
Let $A_l$ be the set of matrices satisfying (M1), (M2), (M3) and
(M4) such that
\begin{equation}\label{e:RndMtx_WkSelRul_12}
\sup_{j\in\Gamma^{\sharp c}}
\abs{\ip{\phi_j(\omega)}{\mathbf{u}_{l-1}}}<\frac{\alpha}{2\sqrt{k}}.
\end{equation}
If $\mathbf{\Phi}\in A_l$ we have
\begin{eqnarray*}
  \frac{\norm{\mathbf{\Phi}^\ast_{\Gamma^{\sharp c}}(\omega)\mathbf{r}^{l-1}}_\infty}
        {\norm{\mathbf{\Phi}^\ast_{\Gamma^\sharp}(\omega)\mathbf{r}^{l-1}}_\infty}
    &\leq& \frac{\sqrt{k}\norm{\mathbf{\Phi}^\ast_{\Gamma^{\sharp c}}(\omega)\mathbf{r}^{l-1}}_\infty}
        {\norm{\mathbf{\Phi}^\ast_{\Gamma^\sharp}(\omega)\mathbf{r}^{l-1}}_2}
        = 2\sqrt{k}\norm{\mathbf{\Phi}^\ast_{\Gamma^{\sharp c}}(\omega)\mathbf{u}_{l-1}}_\infty \\
    &=& 2\sqrt{k}\sup_{j\in\Gamma^{\sharp c}}\abs{\phi_j(\omega),\mathbf{u}_{l-1}} <\alpha,
\end{eqnarray*}
and this is enough to assure that $\mathcal{I}_n(\alpha)\subset
\Gamma^\sharp$, this is, $\mathbf{\Phi}\in E_l$. Therefore,
\begin{equation}\label{e:RndMtx_WkSelRul_13}
\Prb{E_l|\cap_{n=1}^{l-1}E_n}\geq \Prb{A_l|\cap_{n=1}^{l-1}E_n}.
\end{equation}
Let $B_l$ be the set of matrices $\mathbf{\Phi}(\omega)$ satisfying
(M1), (M2), (M3) and (M4) such that
\begin{equation}\label{e:RndMtx_WkSelRul_14}
\norm{\mathbf{\Phi}^\ast_{\Gamma^\sharp}(\omega)\mathbf{r}^{n-1}}_2\geq
\frac{1}{2}\norm{\mathbf{r}^{n-1}}_2.
\end{equation}
From (\ref{e:RndMtx_WkSelRul_13}) we deduce
\begin{equation}\label{e:RndMtx_WkSelRul_15}
\Prb{E_l|\cap_{n=1}^{l-1}E_n}\geq\Prb{A_l\cap
    B_l|\cap_{n=1}^{l-1}E_n} = \Prb{A_l|B_l\cap(\cap_{n=1}^{l-1}E_n)}
        \Prb{B_l|\cap_{n=1}^{l-1}E_n}.
\end{equation}
Conditioned to $B_l$,
$$\norm{\mathbf{u}_{l-1}}_2=\frac{1}{2}\frac{\norm{\mathbf{r}^{l-1}}_2}{\norm{\mathbf{\Phi}^\ast_{\Gamma^\sharp}\mathbf{r}^{l-1}}_2}\leq 1$$
due to (\ref{e:RndMtx_WkSelRul_14}). By (M1) the columns $\phi_j$ of
$\mathbf{\Phi}(\omega)$ are independents among them. Besides,
conditioned to $\cap_{n=1}^{l-1}E_n$, the vector $\mathbf{r}^{n-1}$
only depends on the columns $\phi_j$ with $j\in\Gamma^\sharp$ since
$\mathcal{I}_n(\alpha)\subset\Gamma^\sharp$, $n=1,2,\ldots,l-1$.
Hence, $\mathbf{u}_{l-1}$ is independent from the columns $\phi_j$
with $j\in\Gamma^{\sharp c}$ and we can use (M3). Then,
\begin{eqnarray}
\nonumber
  \Prb{A_l|B_l\cap(\cap_{n=1}^{l-1}E_n)}
    &=& \Prb{\sup_{j\in\Gamma^{\sharp c}}\abs{\ip{\phi_j(\omega)}{\mathbf{u}_{l-1}}}<\frac{\alpha}{2\sqrt{k}}|B_l\cap(\cap_{n=1}^{l-1}E_n)} \\
\nonumber
    &=& \prod_{j\in\Gamma^{\sharp c}} \Prb{\abs{\ip{\phi_j(\omega)}{\mathbf{u}_{l-1}}}<\frac{\alpha}{2\sqrt{k}}|B_l\cap(\cap_{n=1}^{l-1}E_n)} \\
    &\geq& (1-q_1e^{-c_1\frac{\alpha^2}{4k}m})^{N-k} \geq
    1-q_1(N-k)e^{-c_1\frac{\alpha^2}{4k}m}.
\end{eqnarray}
Conditioned to $\cap_{n=1}^{l-1}E_n$, it is not hard to prove that
the vector $\mathbf{r}^{n-1}\in\text{span
}(\mathbf{\Phi}_{\Gamma^\sharp})$. We can use (M4) to get
\begin{equation}\label{e:RndMtx_WkSelRul_16}
\Prb{B_l|\cap_{n=1}^{l-1}E_n}\geq 1-q_1D^ke^{-c_2 m}\geq1-q_1e^{-c_2
m/2}
\end{equation}
taking $-c_2 m+k\ln D\leq -c_2\frac{m}{2}$, this is,
$m\geq\frac{2k}{c_2}\ln D$.

Substituting (\ref{e:RndMtx_WkSelRul_15}) y
(\ref{e:RndMtx_WkSelRul_16}) en (\ref{e:RndMtx_WkSelRul_14}) and
proceed as in the calculations that lead to
(\ref{e:RndMtx_WkSelRul_9}) to get
\begin{equation}\label{e:RndMtx_WkSelRul_17}
\Prb{E_l|\cap_{n=1}^{l-1}E_n}\geq
1-q_3(N-k)e^{-c_1\frac{\alpha^2}{4k}m},
\end{equation}
with $q_3=q_1+q_2$. Substitute (\ref{e:RndMtx_WkSelRul_17}) and
(\ref{e:RndMtx_WkSelRul_10}) in (\ref{e:RndMtx_WkSelRul_11}) to get
\begin{eqnarray*}
  \Prb{\cap_{n=1}^l E_n}
    &\geq& (1-q_3(N-k)e^{-c_1\frac{\alpha^2}{4k}m})
        (1-q_3(l-1)(N-k)e^{-c_1\frac{\alpha^2}{4k}m}) \\
    &\geq& 1-q_3(N-k)e^{-c_1\frac{\alpha^2}{4k}m}-q_3(l-1)(N-k)e^{-c_1\frac{\alpha^2}{4k}m} \\
    &=& 1-q_3l(N-k)e^{-c_1\frac{\alpha^2}{4k}m},
\end{eqnarray*}
which is what we wanted to prove. For the last probability be
greater than $0$ we must take
$$m\geq \frac{4k}{c_1\alpha^2}\ln q_3l(N-k).$$
\hfill $\blacksquare$ \vskip .5cm   

We can now state similar comments to those at the end of the proof
of Theorem \ref{t:RndMtx_RlxWkSelRul}. We emphasize next corollary,
that in the case $\alpha=1$ gives Theorem 6 of \cite{TrGi}.
\begin{cor}\label{c:t:RndMtx_WkSelRul}
Choose the same conditions as in Theorem \ref{t:RndMtx_WkSelRul},
substituting (\ref{e:RndMtx_WkSelRul}) by
\begin{equation}\label{e:c:t:RndMtx_WkSelRul}
m\geq C\frac{k}{\alpha^2}\ln (q_3N^2/4)
\end{equation}
with $C$ large enough. Then, WOMP algorithm recovers the $k$-sparse
vector $\mathbf{x}$ in the first $k$ iterations with probability
greater than or equal to
$$1-q_3\frac{N^2}{4}e^{-c_1\frac{\alpha^2}{4k}m}.$$
\end{cor}


\section{Some Experiments.}\label{Sec:Exprmnts} In this section we
present experiments on the sparse recovery and non sparse
approximation problems for the orthogonal and gradient algorithms
with the selection rule
$\mathcal{\tilde{I}}=\mathcal{\tilde{I}}(\alpha)$ given by
(\ref{e:Sel_Rule_tilde}) and compare them against the results
obtained with the more classical selection rule
$\mathcal{I}=\mathcal{I}(\alpha)$ given by (\ref{e:Sel_Rule}). In
every set of experiments one Gaussian matrix was created of order
$m\times N$. For some signals the results are shown as percentage of
elements recovered and for others we use the \textbf{signal-to-noise
ratio} of the energy of the original signal
$\mathbf{x}\in\mathbb{R}^N$ and the energy of the difference between
the signal and the approximation $\mathbf{a}$, given by
$$SNR=10\log_{10}(\frac{\norm{\mathbf{x}}_2}{\norm{\mathbf{x}-\mathbf{a}}_2}).$$

\begin{figure}[!t]
\centering \includegraphics[width=5in]{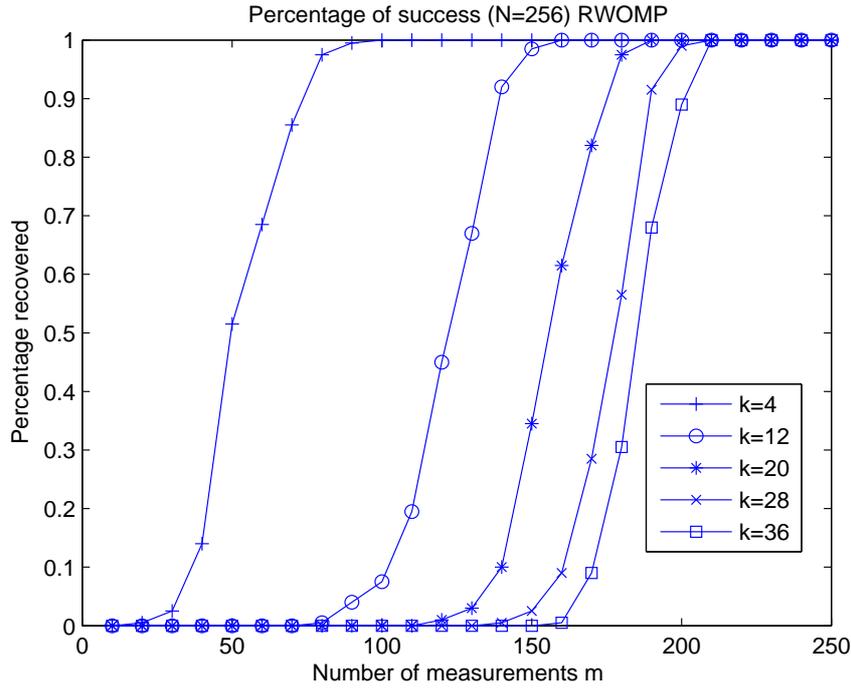} \caption{Recovery
of sparse signals with RWOMP, $\alpha=0.125$. For each set $(m,k)$,
200 experiments were generated with one Gaussian matrix of order
$m\times 256$. Up to $k$ iterations are
allowed.}\label{f:SPRS_RWOMP}
\end{figure}

Figure \ref{f:SPRS_RWOMP} shows the percentage of elements recovered
with the RWOMP algorithm with parameter $\alpha=0.125$. The Gaussian
matrices generated have $N=256$ and $m=10\ell$,
$\ell=1,2,\ldots,25$. The sparsity levels have been chosen to be
$k=4,12,20,28,36$, (each graph corresponds to one of them). For each
pair $(m,k)$, 200 experiments were run for different signals.
The results can be compared with those obtained on Figure 1 of
\cite{TrGi} for OMP ($\alpha=1$). The parameters $N,k$ and $m$ take
the same value, but 1000 experiments for each set were performed in
\cite{TrGi}. The results for RWOMP are better that those for OMP in
\cite{TrGi} for $k=20,28,36$.

\begin{figure}
\centering\includegraphics[width=5in]{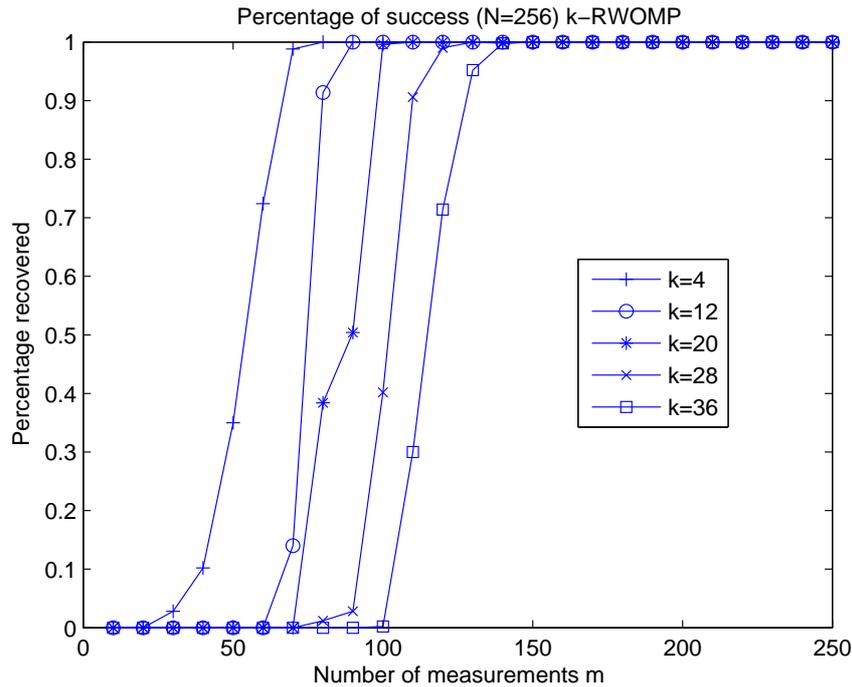}
\caption{Recovery of sparse signals with $k-$RWOMP, with the same
parameters as in Figure \ref{f:SPRS_RWOMP}. The results are better
than in Figure \ref{f:SPRS_RWOMP} for RWOMP.  Up to $k$ iterations
are allowed.}\label{f:SPRS_kRWOMP}
\end{figure}

Next, for computational purposes, we introduce a minor modification
of the RWOMP algorithm, called $k$-RWOMP algorithm. At each
iteration in RWOMP we keep the $k$-largest elements of the
orthogonal approximation $\mathbf{x}^n$. Figure \ref{f:SPRS_kRWOMP}
shows the results of applying the $k$-RWOMP with the same parameters
of the experiments described for Figure \ref{f:SPRS_RWOMP}. As can
be seen, in Figure \ref{f:SPRS_kRWOMP} exact recovery is achieved
with smaller values of $m$.

\

The next set of experiments is done on images. We take an image of
$64\times 64$ pixels, which is part of Lena. Two decompositions of
the wavelet transform were performed using \emph{Daubechies 5}
wavelets. The compressed sensing was done only on the detail
coefficients. The sparsity $k$ is calculated as the integer part of
$5\%$ of $L$, where $L$ is the number of vertical, horizontal or
diagonal coefficients at each decomposition level. The number of
measurements at each level is the integer part of $k\log_2(L/k)$ for
vertical, horizontal and diagonal coefficients. The k-RWOMP and
k-RWGP algorithms (those with selection rule
$\tilde{\mathcal{I}}(\alpha)$ as defined in
(\ref{e:Sel_Rule_tilde})) were run with $\tilde\alpha=0.125$ and
$\tilde\alpha=0.15$, respectively. The WGP algorithm (with selection rule
$\mathcal{I}(\alpha)$ as defined in (\ref{e:Sel_Rule})) was run with
$\alpha=0.8$. The relevant data is given in Figure 3.
When the sparsity goes to
$10\%$ of $L$ it takes several minutes for the CoSaMP algorithm to
 stop, whereas for the rest of algorithms it takes almost the same
time. For each algorithm, the running times and the $SNR$'s differ
for different $\alpha$'s and, moreover, they may even differ with
the same $\alpha$'s due to the randomness of $\mathbf{\Phi}$. Also
visually there is an improvement for the k-Relaxed Weak algorithms
for which there are less artifacts in the smoother regions of the
image. Judging by the SNR, k-RWGP gives the best approximation.

\begin{figure}
\setlength{\unitlength}{10pt}
\begin{center}
\includegraphics[width=6in]{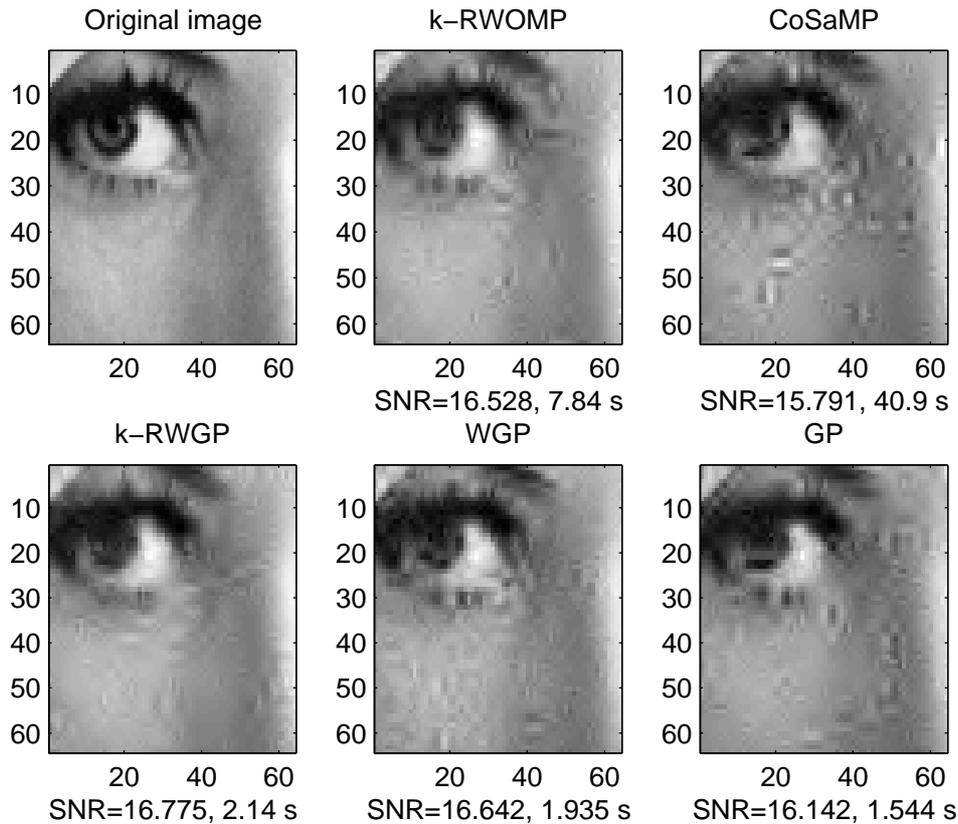}
\end{center}
\caption{k-RWOMP and k-RWGP vs CoSaMP, SWGP and GP.
}\label{f:IMAG2x3}
\end{figure}



\begin{thebibliography}{99}

\parskip=0.1cm

\bibitem{Ach}
D. Achlioptas, \emph{Database-friendly Random Projections:
Johnson-Lindenstrauss with Binary Coins}, Journal of Comp. \& Sys.
Sci., 66 (4), (2003), p.671-687, special issue of invited papers
from PODS'01.

\bibitem{Bar93}
A.R. Barron, ``Universal approximation bounds for superposition of $n$ sigmoidal functions",
 \emph{IEEE Trans. Inform. Thoery}, vol. 39,
 pp. 930-945, 1993.

\bibitem{BDDW}
R. Baraniuk, M. Davenport, R. DeVore and M. Wakin, \emph{A Simple
Proof of the Restricted Isometry Property for Random Matrices},
Constructive Approximation, vol. 28, no. 3, pp. 253-263, 2008.

\bibitem{BlDa1}
T. Blumensath, M. E. Davies; ``Gradient Pursuits", \emph{IEEE
Transactions on Signal Processing,} vol. 56, num. 6, pp. 2370-2382,
2008.

\bibitem{BlDa2}
T. Blumensath, M. E. Davies, ``Stagewise Weak Gradient Pursuits",
\emph{IEEE Transactions on Signal Processing} vol. 57, num. 11, pp.
4333-4346, 2009.

\bibitem{CanTao2}\label{bibl_CanTao2}
E. Candès and T. Tao, ``Decoding by linear programming," \emph{IEEE
Trans. on Information Theory,} vol. 51, num. 12, pp. 4203-4215,
2005.

\bibitem{CanTao1}\label{bibl_CanTao1}
E. Candès and T. Tao, ``Near optimal signal recovery from random
projections: Universal encoding strategies?" \emph{IEEE Trans. on
Information Theory,} vol. 52, num. 12, pp. 5406-5425, 2006.

\bibitem{CoDaDe2}\label{bibl_CoDaDe2}
A. Cohen, W. Dahmen, R.A. DeVore, ``Instance Optimal Decoding by
Thersholding in Compressed Sensing," \emph{Proceedings of the 8th
International Conference in Harmonic Analysis and Partial
Differential Equations, El Escorial 2008}, Contemporary Mathematics,
AMS, 2009.

\bibitem{DM09}
W. Dai, O. Milenkovic, ``Subspace pursuit for compressive sensing: closing the gap between performance and complexity",
 \emph{IEEE Trans, Inform, Theory}, vol. 55, no 5,
 pp. 2230-2249, 2009.

\bibitem{DasGup}
S. Dasgupta and A. Gupta, \emph{An elementary proof of the
Johnson-Lindenstrauss lemma}, ICSI Technical Report TR-99-006, U.C.
Berkeley, 1999.

\bibitem{DW10}
M. Davenport, M. Watkin, ``Analysis of Orthogonal Matching Pursuit
using the Restricted Isometry Property",
 \emph{IEEE Trans, Inform, Theory}, vol. 56, no 9,
 pp. 4395-4401, 2010.

\bibitem{DT96}
R.A. DeVore, V.N, Temlyakov, ``Some remarks on greedy algoritms",
 \emph{Adv. in Comput. Math.}, vol. 5,
 pp. 173-187, 1996.

\bibitem{DeV2}
R. A. DeVore, \emph{Deterministic construction of compressed sensing
matrices}, Journal of Complexity, 23:918-925, 2007.

\bibitem{Don}\label{bibl_Don}
D.L. Donoho, ``Compressed sensing", \emph{IEEE Trans. on Information
Theory,} vol. 52, num. 4, pp. 1289-1306, 2006.

\bibitem{DTDS} D.L. Donoho, Y. Tsaig, I. Drori,
J-L. Starck, ``Sparse Solution of Underdetermined Linear Equations
by Stagewise Orthogonal Matching Pursuit", \emph{IEEE Trans. on
Information Theory}, vol. 58, num. 2, pp. 1094-1121, 2012.

\bibitem{GrVa}
R. Gribonval, P. Vandergheynst, ``On the exponential convergence of
Matching Pursuits in Quasi-Incoherent Dictionaries,"  \emph{IEEE
Trans. Information Theory,} vol. 52, num. 1, pp. 255-261, 2006.


\bibitem{Jon87} L. Jones, ``On a conjecture of Huber concerning the convergence of
projection pursuit regression", \emph{The Annals of Statistics},
vol. 15, pp. 880-882, 1987.

\bibitem{LT}
E. Liu, V. N. Temlyakov, ``Orthogonal supergreedy algorithm and
applications in Compressed Sensing", Preprint.

\bibitem{Lu}
G. Lugosi. ``Concentration-of-measure inequalities. Lecture Notes,
december 2004," [On-line]. Available on
\emph{http://www.econ.upf.edu/~lugosi/surveys.html}

\bibitem{Maleh}
R. Maleh, ``Improved RIP Analysis of Orthogonal Matching Pursuit",
Preprint.

\bibitem{Mal}
S. Mallat. \emph{A Wavelet tour on signal processing}, Academic
Press, 2nd. ed., 1999.

\bibitem{MS} Q. Mo, Y. Shen, ``Remarks on the Restricted Isometry Property in Orthogonal Matching Pursuit
algorithm", Preprint, 2012.

\bibitem{MaZh93} S. Mallat, Z. Zhang, ``Matching Pursusit with time-frequency
dictionaries", \emph{IEEE Trans. Signal Processing}, vol. 41, num.
12, pp. 3397-3415, 1993.

\bibitem{NeVe1}
D. Needell, R. Vershynin, ``Uniform uncertainty principle and signal
recovery via regularized orthogonal matching pursuit," \emph{Found.
Comput. Math.}, vol. 9, pp. 317-334, 2009.

\bibitem{NeTr}
D. Needell, J. A. Tropp, ``CoSaMP: Iterative signal recovery from
incomplete and inaccurate samples," \emph{Appl. Comput. Harmon.
Anal.}, vol. 26, num. 3, pp. 301-321, 2009.


\bibitem{RuVe1}\label{bibl_RuVe1} M. Rudelson,
R. Vershynin, ``Sparse reconstruction by convex relaxation: Fourier
and Gaussian measurements," \emph{CISS 2006 (40th Annual Conference
on Information Sciences and Systems)}.

\bibitem{Tem00}
V.N. Temlyakov, ``Weak greedy algorithms",  \emph{Adv. in Comput. Math.}, vol. 12,
 pp. 213-227, 2000.

\bibitem{Tem03}
V.N. Temlyakov, ``Nonlinear methods of approximation",  \emph{Found. Comput. Math.}, vol. 17,
 pp. 269-280, 2003.

\bibitem{Tem11}
V.N. Temlyakov, ``Greedy approximation",  \emph{Cambridge Monographs on Applied and Computational Mathematics},
Cambtidge University Press, 2011.

\bibitem{Tr1}
J. A. Tropp, ``Greed is good: Algorithmic results for sparse
approximation," \emph{IEEE Trans. Info. Theory}, vol. 50, num. 10,
pp. 2231-2242, 2004.

\bibitem{TrGi}
J. A. Tropp, A. Gilbert, ``Signal recovery from random measurements
via Orthogonal Matching Pursuit," \emph{IEEE Trans. Info. Theory},
vol. 53, num. 12, pp. 4655-4666, 2007.

\end{thebibliography}
\end{document}